\documentclass[12pt]{article}
\usepackage[utf8]{inputenc}
\usepackage[english]{babel}
\usepackage{csquotes}
\usepackage{mathtools} 
\usepackage{amssymb}
\usepackage{amsthm}
\usepackage{bm}
\usepackage{graphicx,caption,subcaption}
\usepackage{xcolor}
\usepackage{enumitem}
\usepackage{listings}
\usepackage{setspace}
\usepackage[euler]{textgreek}
\usepackage{hyperref}
\usepackage[top=1.3in, bottom=1.5in, left=1in, right=1in]{geometry}
\usepackage[backend=biber, style=nature]{biblatex}
\usepackage{afterpage}
\newcommand\myemptypage{
    \null
    \thispagestyle{empty}
    \addtocounter{page}{-1}
    \newpage
    }
\newcommand\yemptypage{
    \null
    \thispagestyle{plain}
    \newpage
    }

\newcommand{\sgn}{
\operatorname{sgn}
}

\newcommand{\Tr}{
\operatorname{Tr}
}

\newcommand{\boxtext}[1]{\vspace{0.3cm}

\framebox{\centering
\begin{minipage}{0.97 \textwidth}

#1
\end{minipage}}
\vspace{0.3cm}

}

\newtheoremstyle{mystyle}
  {}
  {}
  {\normalfont}
  {}
  {\bf}
  {.}
  { }
  {\thmname{#1}\thmnumber{ #2}\thmnote{\textbf{\,: #3}}}

\theoremstyle{mystyle}

\newtheorem{defi}{Definition}[section]

\newtheorem{ex}{Example}[section]

\begin{document}
\fontsize{13}{15.6}\selectfont
\begin{titlepage}
	\centering
	\includegraphics[width=0.15\textwidth]{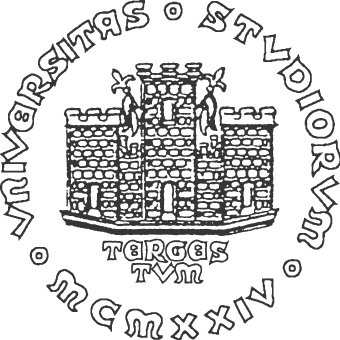}\par\vspace{1cm}
	{\scshape\LARGE Università di Trieste \par}
	\vspace{1.5cm}
	{\scshape\Large A.A. 2021/2022\par}
	\vspace{3cm}
	{\huge\bfseries On the capacity of neural networks \par}
	\vspace{2cm}
	{\Large\scshape Leonardo Cruciani \par}
	\vfill
	{\large Thesis Supervisor: \par
	Prof.~F. Benatti}
\end{titlepage}
\newpage
\begin{titlepage}
	\centering
	\includegraphics[width=0.15\textwidth]{IMG/logo.png}\par\vspace{1cm}
	{\scshape\LARGE Università di Trieste \par}
	\vspace{1.5cm}
	{\scshape\Large A.A. 2021/2022\par}
	\vspace{3cm}
	{\huge\bfseries Sulla capacità delle reti neurali \par}
	\vspace{2cm}
	{\Large\scshape Leonardo Cruciani \par}
	\vfill
	{\large Relatore: \par
	Prof.~F. Benatti}
\end{titlepage}
\myemptypage

 \pagenumbering{roman} 
 \section*{Abstract in English}
The aim of this thesis is to compare the capacity of different models of neural networks. We start by
analysing the problem solving capacity of a single perceptron using a simple combinatorial argument. After
some observations on the storage capacity of a basic network, known as an associative memory, we introduce
a powerful statistical mechanical approach to calculate its capacity in the training rule-dependent Hopfield
model. With the aim of finding a more general definition that can be applied even to quantum neural nets,
we then follow Gardner’s work, which let us get rid of the dependency on the training rule, and comment
the results obtained by Lewenstein et al. by applying Gardner’s methods on a recently proposed quantum
perceptron model.

\section*{Abstract in Italiano}
Lo scopo di questa tesi è comparare la capacità di diversi modelli di reti neurali.
Tramite un semplice calcolo combinatiorio, viene valutata la capacità del \textit{perceptron} semplice. Dopo alcune osservazioni riguardo alla capacità di un elementare modello di rete, chiamato memoria associativa, viene introdotto l'approccio meccanico-statistico, che permette il calcolo della capacità della rete in un modello, detto di Hopfield, dipendente dalla regola di apprendimento dei neuroni. Dunque, con l'obiettivo di trovare un approccio che permetta di svincolarsi da questa dipendenza, vengono seguiti i lavori di E. Gardner sul calcolo della capacità.
Infine vengono commentati i risultati ottenuti da Lewenstein et al. applicando l'approccio di Gardner ad un modello di neurone quantistico recentemente proposto.
 \newpage
\yemptypage
 \newpage
\vspace*{\fill}
\begin{displayquote}[Catullus, \textit{Carmen LXVI}\footnote{Elegiac couplets from the Latin translation by Catullus of \textit{The Lock of Berenice}. The poem, arrived incomplete, was written by the Hellenistic poet Callimachus to celebrate Berenice, the beautiful queen of Egypt.
Her lock, consecrated in a vow and then disappeared, was found by the court astronomer Conon in the firmament, as a constellation.\cite{Catullo}}]
\centering
Omnia qui magni dispexit lumina mundi,\\
qui stellarum ortus comperit atque obitus,\\
flammeus ut rapidi solis nitor obscuretur,\\
ut cedant certis sidera temporibus\\
ut Triviam furtim sub Latmia saxa relegans\\
dulcis amor gyro devocet aereo:\\
idem me ille Conon caelesti in lumine vidit\\
e Beroniceo vertice caesariem\\
fulgentem clare, quam multis illa dearum\\
levia protendens brachia pollicita est,\\
qua rex tempestate novo auctus hymenaeo\\
vastatum finis iverat Assyrios,\\
dulcia nocturnae portans vestigia rixae,\\
quam de virgineis gesserat exuviis.\\
\end{displayquote}
\vspace*{\fill}

 \newpage
\yemptypage

\tableofcontents

 \newpage

\pagenumbering{arabic} 
 
 \section{Introduction}
 Machine learning and quantum computing are two technologies that each have the potential to change the way computing works to solve previously intractable problems. \\
 Machine learning methods are ubiquitous in pattern recognition and image classification, dynamical system modeling and forecasting, and in general for all kind of task that are difficult or just impractical to hardcode.
However there are limitations to successfully solving such classification problems when the feature space is large and the functions used in its implementation are computationally expensive to evaluate. \\ 
On their own Quantum Information Theory and Quantum Computing have been making great leaps forward in the last years, both technologically and theoretically. Their applications now range from physics~\cite{Physics} and chemistry~\cite{Chemistry} simulations to cryptography~\cite{Cryptography}, finance~\cite{Finance} and industrial optimization~\cite{Industry}.
Particularly, they proved to be extremely good at dealing with some kind of classically time-consuming problems. For example, the advantages brought by Quantum Computing for tasks such as searching an unsorted database~\cite{Grover}, the factorization of the product of two large prime numbers~\cite{Shor} and other NP-complete problems, often requiring massively parallel computing if solved on classical hardware, are well renown.\\
For its nature, the training of a neural net may be one of the problems where Quantum Computing can really show its true potential. 
But the benefits of quantum machine learning in the short term are not clear, and even if  some models for quantum neural networks have been recently proposed \cite{Benatti, Petruccione}, understanding the structure and training of quantum models, and quantum neural networks in particular, requires further research~\cite{Abbas}.\\
We may start by asking, what do we call a \textit{quantum neural network}?
Should it be possible to implement it only on quantum hardware or should the algorithm show at least some advantages when implemented on quantum hardware? How can we measure such advantages? 
To answer these questions, we try to compare the \textit{problem solving} capacity of the mathematical models behind traditional, classical neural networks with an alternative, \textit{quadratic} model of neuron, which have been implemented~\cite{Macchiavello} on a quantum circuit. \\
We start by analysing the capacity of a single perceptron: we discuss its limitations, various solutions - which let us introduce multi-perceptron networks and perceptrons with different activation functions, and directly calculate the capacity of the simple perceptron using a combinatorial argument. We then make some remarks about measuring its complexity in actual implementations.\\
After some observations on the storage capacity of a basic network, called an associative memory, we introduce a powerful statistical mechanical approach to calculate its capacity in the training rule-dependent Hopfield model.   
With the aim of finding a learning-phase independent model to calculate the capacity, so that it can be applied even to the \textit{quadratic} quantum neural net, we follow Gardner's~\cite{SantaFe} work, and comment the results obtained by Lewenstein et al.~\cite{Lewenstein} by applying Gardner's methods on the recently proposed quantum perceptron model~\cite{Macchiavello}.

 \newpage
\section{The perceptron}
\begin{defi}[Perceptron]
As defined by McCulloch and Pitts \cite{mcculloch}, the perceptron is a function that maps its input $\bm \xi$, usually a vector of binary values, to an output value $\bm \xi \mapsto \zeta$, usually a binary value itself. In order to do so, it calculates the dot product between the input vector and a vector of (usually) real values $\bm w$, called \textbf{weights}, adds a threshold $\theta$ and computes a non-linear function, like the Heaviside $\Theta$ or the sign function $\sgn$.
In this thesis we will mainly consider the non-linear function:
\begin{equation}
\zeta = \sgn(\bm w \cdot \bm \xi-\theta).
\end{equation}
\end{defi}
The perceptron is the fundamental unit of artificial networks, and can be seen as both a \textbf{binary classifier}, namely a function which can decide whether or not an input belongs to some specific class, and as the elemental \textbf{storage unit} of a memory. As we will see, these concepts are strictly related, and in both cases it is useful to define the concept of \textbf{capacity}. In the first case, it is the number of problems that the perceptron can solve, in the second, it measures the amount of information that can be stored in a perceptron.\\
Let us start with the concept of capacity as a measure of what a perceptron can and cannot do - perceptrons in fact are not universal computers, since they can compute the AND and OR boolean functions, but not the XOR, as we will see soon.
\begin{figure}[h!]
\vspace{1cm}
\centering
\includegraphics[width=0.6\linewidth]{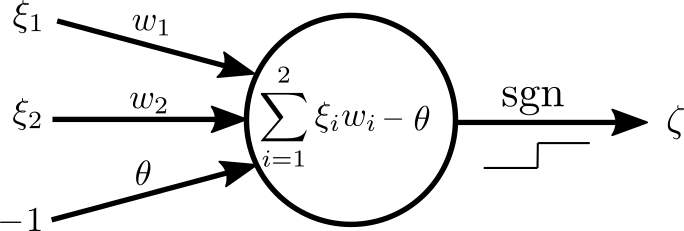}
\vspace{0.5cm}
\caption{Scheme of a single perceptron with 2 inputs $\xi_1, \xi_2$, 2 weights $w_1,w_2$ and a threshold $\theta$, implementing a $\sgn$ activation function.}
\label{simp}
\end{figure}

 \subsection{Limitations of a single perceptron}
Let us consider a single perceptron with $N$ binary inputs $\xi_i=\pm 1$ and one binary output $\zeta$.  Collectively, the $N$ input values form a pattern $\bm \xi=( \xi_1, \hdots, \xi_N)$.
We call an \textbf{example} the couple \textit{(given input pattern, desired output)}. Given an ordered set of $p \leq 2^N$ examples, we thus identify the $\mu$-th example with the couple $(\bm \xi^\mu, \zeta^\mu)$. Such a set constitutes a \textbf{$\bm p\,$-example problem}.\\
We say the perceptron can solve exactly a $p$-example problem if, given the pattern $\bm \xi^\mu$ as input, the perceptron outputs the desired value $\zeta^\mu$, for all $\mu=1, \hdots, p$. Since $\zeta$ is a binary output, from $p$ examples we can define $2^p$ different problems. The perceptron will be able to solve only a few of these, i.e. for the most part of these problems it does not exist a configuration of the weights such that all the associations $\bm \xi^\mu \to \zeta^\mu$ are symultaneously satisfied.
\begin{ex}[XOR problem] \label{ex1}
This famous~\cite{MP} example shows that a perceptron with 2 binary inputs $\xi_1,\xi_2$ cannot implement the exclusive disjunction XOR. Here is the truth table of the binary operator: 
\begin{equation}
\begin{array}{rrr}
\xi_1&\xi_2&XOR\\\hline
-1&-1&-1\\
-1&1&1\\
1&-1&1\\
1&1&-1\\
\end{array}
\end{equation}
The $4$-example problem we would like to solve is exactly the implementation of this truth table. Let us call $w_1$ and $w_2$ the weights associated to the inputs $\xi_1,\xi_2$ and $\theta$ the threshold, so that:
\begin{equation}
\zeta = \sgn(w_1\xi_1 +w_2\xi_2-\theta).
\end{equation}
The values of  $w_1, w_2,\theta$ must then satisfy:
\begin{equation}
\begin{cases}
-w_1-w_2-\theta <0,\\
-w_1+w_2-\theta > 0,\\
w_1-w_2-\theta > 0,\\
w_1+w_2-\theta <0,\\
\end{cases}
\end{equation}
but this system does not admit any solution, as summing the first and the last equation gives $\theta > 0$, while summing the central ones gives $\theta < 0$. It is usually said that the XOR values are not \textbf{linearly separable} (for a graphical representation, see fig. \ref{sep2}).
\end{ex}
\begin{figure}[h!]
\vspace{1cm}
\centering
\includegraphics[width=0.8\linewidth]{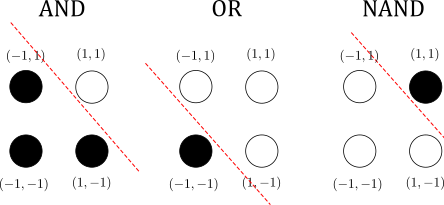}
\vspace{0.5cm}
\caption{White ball means output is $1$, black ball that it is $-1$. Red line shows the linear separability of functions OR, AND and NAND.}
\label{sep}
\end{figure}

 \subsubsection{The deep perceptron}\label{ffnet}
It is however possible to solve the XOR problem by using a configuration of multiple perceptrons, called a \textbf{deep perceptron} or \textbf{feed-forward network}. Here we show how it is done.
\begin{ex}[Solving the XOR problem with a deep perceptron]
We should first recognise that, while unable to implement the XOR, the single perceptron can represent both the OR and the AND (along with their negatives NOR and NAND, see fig. \ref{sep}):
\begin{equation}
\begin{array}{rr|r|r|r}
\xi_1&\xi_2&AND&OR&\sc{N\!AN\!D}\\\hline
-1&-1&-1&-1&1\\
-1&1&-1&1&1\\
1&-1&-1&1&1\\
1&1&1&1&-1\\
\end{array}
\end{equation}

Moreover, it can be easily checked that the XOR function can be implemented by a convolution of AND, OR and NAND functions:
\begin{equation}
\sc{XOR}(\xi_1,\xi_2)=\sc{AND}(\sc{N\!AN\!D}(\xi_1,\xi_2), OR(\xi_1,\xi_2)).
\end{equation}
Therefore, we can combine the 3 simple perceptron to obtain a network that implements the XOR, as in fig. \ref{xor}. In fact, it has even been shown that 2-layer networks are universal for computation. 
\begin{figure}[h!]
\vspace{1cm}
\centering
\includegraphics[width=0.6\linewidth]{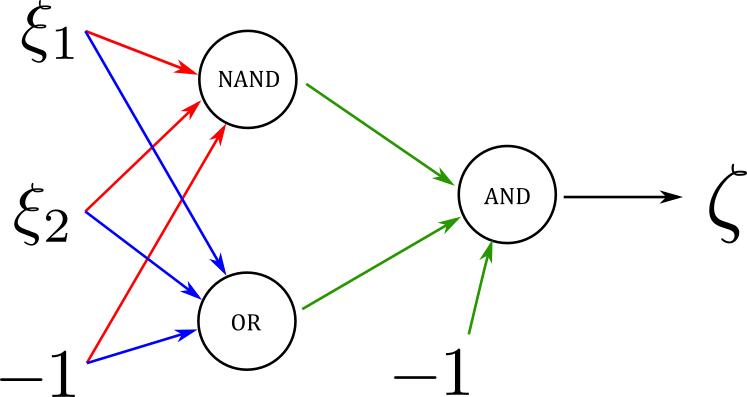}
\vspace{0.5cm}
\caption{Implementation of the XOR boolean operator using a 2-layered neural network.}
\label{xor}
\end{figure}
\end{ex}

\begin{figure}[h!]
\vspace{1cm}
\centering
\includegraphics[width=0.3\linewidth]{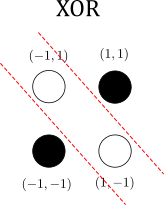}
\vspace{0.5cm}
\caption{One hyperplane is not enough to implement the XOR function, but we can combine an OR and a NAND to do it (compare with fig. \ref{sep})}
\label{sep2}
\end{figure}

\subsubsection{Implementations and complexity}
Both the simple perceptron and the layered network can be easily implemented on an ordinary computer. In fact the complexity of the algorithm scales linearly with the number of inputs, making it $O(N)$.
 
\subsection{Capacity of a simple perceptron} 
\label{percap}
\begin{figure}[h]
\centering
\includegraphics[width=0.7\linewidth]{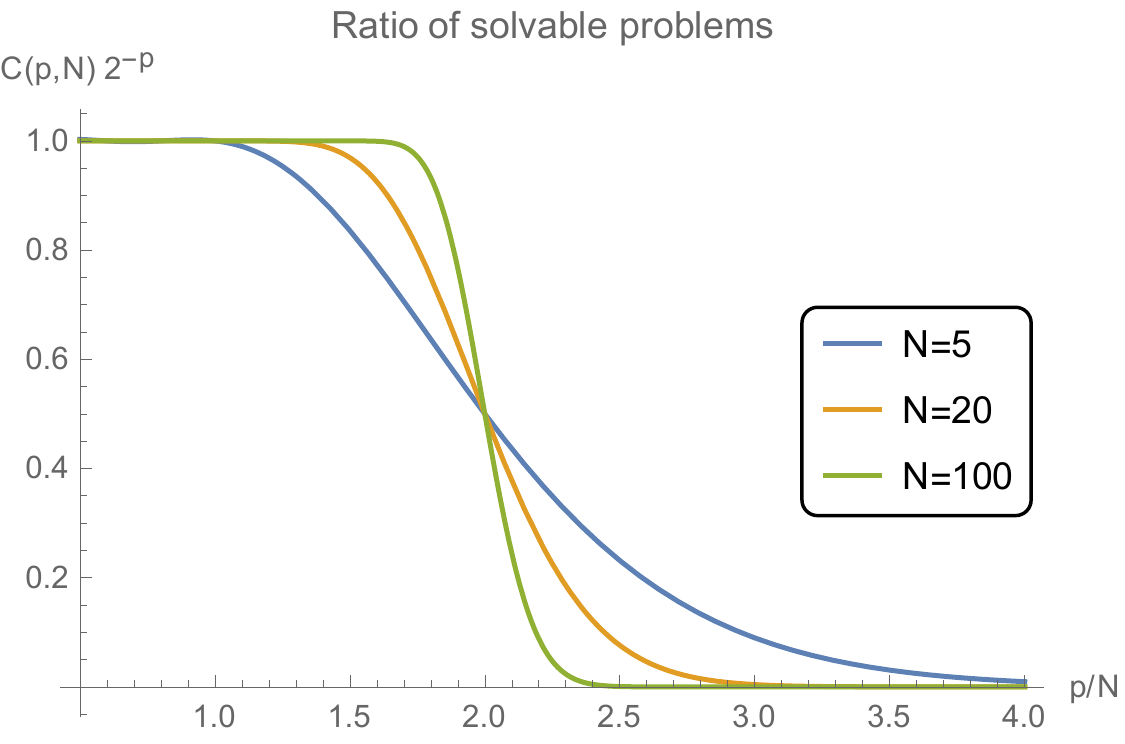}
\caption{The ratio between the number of solvable $p$-problems $C(p,N)$ and the total number of distinct $p$-problems, $2^p$, as a function of the ratio $p/N$.}
\label{capacity}
\end{figure}

$C(p, N)$, the number of $p$-problems that an $N$-input perceptron can solve is a geometrical problem that can be evaluated analytically. 
In fact, the values $\bm \xi^\mu$ that the binary input vector assumes for each example $\mu$ can be associated to one of the vertices of an hypercube in $\mathbb R^N$. The watershed value of the \textbf{net input} $x=\sum_{i=1}^N \xi^\mu_i w_i =0 $ thus defines a subspace passing through the origin.
$C(p,N)$ is then equivalent to the number of ways we can shatter $p$ points in $\mathbb R^N$ using an hyperplane that passes through the origin. 
This value turns out to be (as shown in Appendix \ref{h1}):
\begin{equation} C(p, N) =2\sum^{N-1}_{k=0}\binom{p-1}{k}.\end{equation}
We can see that if $p \leq N$, the sum extends over all the configuration space and the perceptron can solve all of the $2^p$ possible $p$-problems while for $p=2N$ it can only solve $2^{p-1}$ problems, half of the total.
Moreover, as we can see in figure \ref{capacity}, the ratio between solvable and total problems tends to a Heaviside $\Theta$ function in the limit of large $N$.
\begin{equation}
\frac{C(p,N)}{2^p}\xrightarrow{N\to \infty} \Theta(p/N-2).
\end{equation} 
\begin{defi}[Capacity of a perceptron]
The capacity of a simple perceptron is defined as the ratio $\alpha$ between the maximum number $p_*$ of examples such that all the $2^{p_*}$ associated $p_*$-problems are solvable and the number of inputs $N$:
\begin{equation}
\alpha = \lim_{N\to \infty}\frac{p_*}{N}, \qquad p_*=\max\left\{ p \, :\, C(p,N)=2^p\right\}.
\end{equation} 
\end{defi}
For the single perceptron this value is clearly $\alpha = 2$.

\subsection{The \textit{quadratic} perceptron} 
\begin{figure}[h!]
\vspace{0.5cm}
\centering
\includegraphics[width=0.45\linewidth]{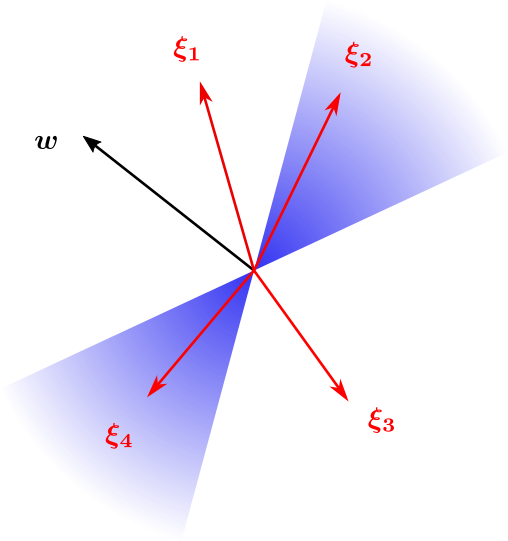}
\caption{The \textit{quadratic} perceptron with weights $\bm w$ and $\kappa = \cos^2(50^\circ)$ classifies examples $\bm \xi_1$ and $\bm \xi_3$ as ``1'', while examples $\bm \xi_2$ and $\bm \xi_4$ are classified as ``0''.}
\label{quad}
\vspace{0.5cm}
\end{figure}

The choice of the non-linear function give us enough room for generalization, for example the sign $\sgn$ function (or equivalently the Heaviside $\Theta$), while useful for theoretical research, are not usually the best choices when it comes to implementing the perceptron (the continuous hyperbolic tangent $\tanh$ or the ramp function \eqref{rampfunc} - often called ReLU - are often preferred \cite{He}).
\begin{equation}\label{rampfunc}
    \operatorname{ramp}(x)= \begin{cases}
   \, x \quad x\geq 0,\\
   \, 0 \quad x < 0.
    \end{cases}
\end{equation}
\begin{figure}[h!]
\vspace{0.5cm}
\centering
\includegraphics[width=0.7\linewidth]{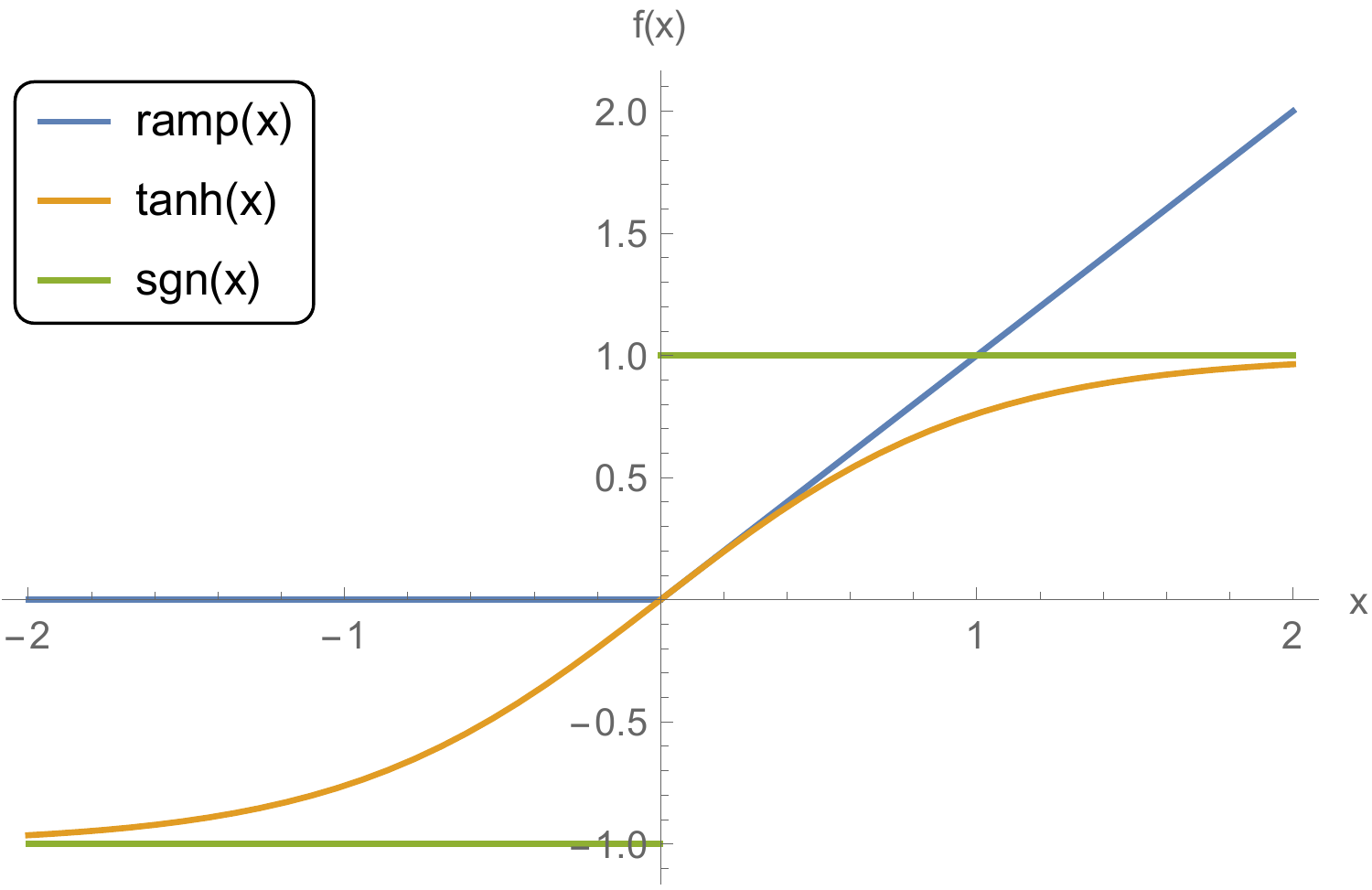}
\caption{Some of the most commonly used activation functions}
\label{transfer}
\vspace{0.5cm}
\end{figure}
With the recent development of quantum computers, new models of perceptron have been proposed, often implementing a quadratic net input to a $\Theta$ function:
\begin{equation}\label{qp}
\zeta = \Theta\left((\bm w \cdot \bm \xi)^2-\kappa\right),
\end{equation}
To understand how this perceptron works, let us first consider the case $\kappa=0$. In this case, we assume $\Theta(0)=0$: the perceptron then classifies as ``0'' the examples $\bm \xi$ that are orthogonal to the weight vector $\bm w$, while all the other values are classified as ``1''. This obviously means that the $0$'s subspace has one dimension less than the example space. We therefore introduce a thereshold $\kappa >0$, which should be small, as we will show in Section \ref{h7}, and we can drop the $\Theta(0)=0$ assumption. An example of classification of some examples $\bm xi_i$ given a weight vector $\bm w$ can be seen in fig. \ref{quad}.
Before introducing the quantum perceptron model developed by Macchiavello et al., let us exemplify the advantage given by the \textit{quadratic} perceptron, and also show that the solution does not depend on $\kappa$, as long as $\kappa>0$.
\begin{ex}[Solving the XOR problem with a \textit{quadratic} perceptron]\label{ex3}
As we did in example \ref{ex1}, we start by writing the system of equations that implement the XOR. For the \textit{quadratic} perceptron, it is:
\begin{equation}
\begin{cases}
(-w_1-w_2)^2-\kappa <0,\\
(-w_1+w_2)^2-\kappa > 0,\\
(w_1-w_2)^2-\kappa > 0,\\
(w_1+w_2)^2-\kappa <0,\\
\end{cases}
\end{equation}
We can immediately notice that the first equation is equivalent to the fourth, and so are the second and the third ones. We are then left with just two equations:
\begin{equation}
\begin{cases}
w_1^2+w_2^2+2w_1w_2-\kappa <0,\\
-w_1^2-w_2^2+2w_1w_2+\kappa < 0,\\
\end{cases}
\end{equation}
 That yields $w_1w_2<0$, ie. $w_1=-w_2$. The problem thus has a solution that does not depend on $\kappa$, as long as $\kappa >0$. The geometrical meaning of $\kappa$ is suggested in fig. \ref{quad2}.
\begin{figure}[!h]
\vspace{0.5cm}
\centering
\includegraphics[width=0.45\linewidth]{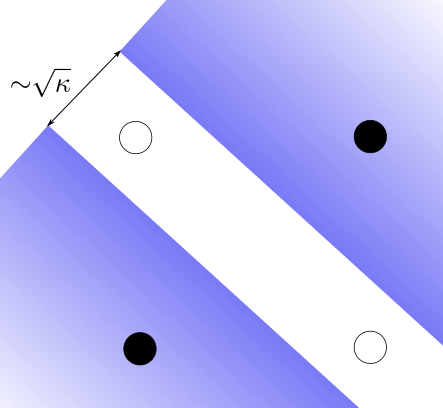}
\caption{A \textit{quadratic} perceptron can single-handedly solve the XOR problem. }
\label{quad2}
\vspace{0.5cm}
\end{figure}
\end{ex}

Looking at fig. \ref{quad2} one may wonder whether this \textit{quadratic} perceptron can implement the other binary functions. The answer is yes, if we add a threshold:
\begin{equation}\label{qp2}
\zeta = \Theta\left((\bm w \cdot \bm \xi-\theta)^2-\kappa\right),
\end{equation}
\begin{ex}[Implementing the NAND in a \textit{quadratic perceptron}] \label{ex4}
To implement the NAND binary function, the following equations should be satisfied:
\begin{equation}
\begin{cases}
(-w_1-w_2-\theta)^2-\kappa >0,\\
(-w_1+w_2-\theta)^2-\kappa > 0,\\
(w_1-w_2-\theta)^2-\kappa > 0,\\
(w_1+w_2-\theta)^2-\kappa < 0,\\
\end{cases}
\end{equation}
Summing the second and the third equations yields:
\begin{equation}
(w_1-w_2)^2+\theta^2-\kappa > 0,
\end{equation}
while subtracting the last from the first yields:
\begin{equation}
(w_1+w_2)\, \theta > 0.
\end{equation}
These equations are clearly satisfied for $w_1=w_2=\theta=1$. Once again the result does not depend on $\kappa \in [0,1]$.\\
For the AND function we have a similar result, but there is a catch. In fact the equations become:
\begin{equation}
\begin{cases}
(w_1-w_2)^2+\theta^2-\kappa < 0,\\
(w_1+w_2)\, \theta < 0.
\end{cases}
\end{equation}
Solved for $w_1=w_2=1$, $\theta<0$ and $\kappa >\theta^2$. Since we will need, however small, $\theta\neq 0$, we will not be able to take the limit $\kappa \to 0$.
\end{ex}

\subsubsection{Complexity} \label{comp2}
As it only implements a $\log N$-digit squaring other than the operations already present in the simple perceptron, therefore the complexity of a \textit{quadratic} perceptron implemented on a classical computer remains  $O(N)$ \cite{harvey}.

\subsubsection{The Macchiavello et al. quantum perceptron}
\begin{figure}[h!]
\centering
\includegraphics[width=0.6\linewidth]{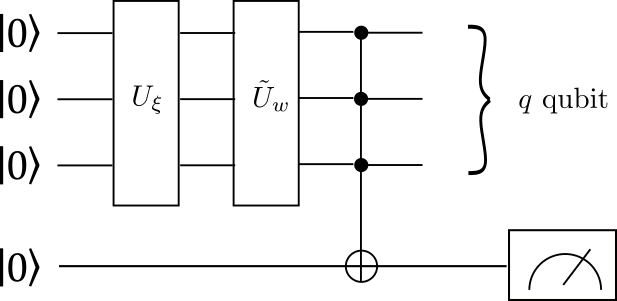}
\caption{The quadratic perceptron implementation in a quantum circuit proposed by Macchiavello et al.\cite{Macchiavello}}
\label{Macchiavello}
\end{figure}

In a recent paper \cite{Macchiavello}, Macchiavello et al. proposed the implementation of a \textit{quadratic} perceptron in a quantum circuit (fig.\ref{Macchiavello}). In particular, it consists in a circuit of $q+1$ qubit (where $q=\lceil \log_2N \rceil$ and $+1$ refers to an \textit{ancilla bit}) implementing the gates $U_\xi$, $\tilde U_w$ and $C^qX$. Given, as an instance, $\bm \xi = (-1,1,1,1)$ and
$\bm w = (-1,-1,1,1)$, the gates $U_\xi$ and $\tilde U_w$ action is the following:
\begin{equation}
    U_\xi |00\rangle = -|00\rangle +|01\rangle +|10\rangle +|11\rangle, \qquad   U_w^\dagger |11\rangle = -|00\rangle -|01\rangle +|10\rangle +|11\rangle,
\end{equation}
while the action of the gate $C^2X$ is:
\begin{equation} \label{CX}
    C^2X\!\!\sum_{j\in \{00,01,10,11\}}\!\!c_j\,|j,0\rangle = \sum_{j\in \{00,01,10\}}\!\!c_j\,|j,0\rangle + c_{11}\,|11,1\rangle.
\end{equation}
The non linear activation function, instead, is implemented with a measurement of the ancilla (see Appendix \ref{h0} for more details). Overall, a $(q+1)$-qubit perceptron works like this:
\begin{enumerate}
\item First, thanks to a procedure called \textit{hypergraph states generation subroutine} (HSGS) - embedded into the $U_\xi$ gate - the binary example vector $\bm \xi$ is saved into the state of the $q$-qubit system.
\item Then, applying the $\tilde U_w$ gate (a gate constructed similarly as $U_\xi$, this time implementing the binary weight vector $\bm w$) to the system evaluates the scalar product $\bm \xi \cdot \bm w$ and saves it into the coefficient of the last ($j=2^q-1$) orthogonal component of the system state, as in \eqref{CX}.
\item Finally, it implements the necessary non-linearity by means of a measurement procedure: 
it applies a gate  $C^qX$ that targets the ancilla qubit, in order to measure it equal to $|1\rangle$ with a probability that is equal to the normalized square modulus of the scalar product between  $\bm \xi$ and $\bm w$, effectively implementing \eqref{qp}.
\end{enumerate}

\newpage

\section{The Hopfield statistical model} 
The simplicity of a single perceptron allows for a combinatorial computation of its geometric capacity. Unfortunately doing so for a network of perceptrons (e.g. an associative memory) or even for a perceptron with a different activation function (e.g. the \textit{quadratic} perceptron, where the net input is: $x=(\bm w \cdot \bm \xi)^2 -\theta$) is much more complicated. One then resort to statistical mechanics to approach such an issue. In this Section we first make some observations about associative memories and their capacity,  we then introduce the energy function and evaluate the capacity with the powerful tools of statistical mechanics.

\begin{figure}[!h]
\centering
\includegraphics[width=0.6\linewidth]{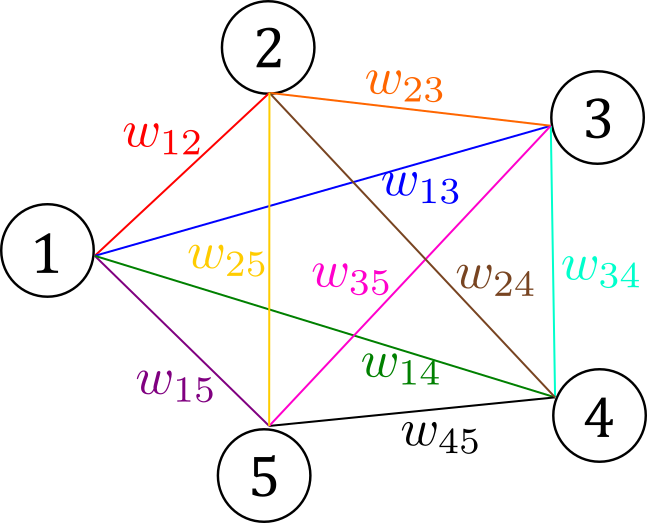}
\caption{A schematic representation of an associative memory with $N=5$ sites.}
\label{net}
\end{figure}

\begin{defi}[Associative Memory]
An associative memory of $N$ units is a network of $N$ binary perceptrons (also called units, or neurons), each taking $N$ binary values $S_i=\pm 1$ as input and outputting a binary value $S'_j$, obeying to the following dynamics:
\begin{equation}\label{dynamics}
S'_i=\sgn\left(\sum_{j=1}^N w_{ij} S_j\right) .
\end{equation}
The weights $w_{ij}$ are real numbers that symbolize the connection between two neurons (see fig. \ref{net}) , and are adjusted during the training of the network in order to have some given stable states.
\end{defi}
\subsection{Pattern Memorization} \label{patmem} Concretely, we want the network to be able to memorize a pattern of binary values \\$(S_1, \hdots, S_N)=(\xi_1, \hdots, \xi_N)=\bm \xi$. The condition for this pattern to be \textit{memorized} is it being a stable point of the dynamics:
\begin{equation}
\sgn\left(\sum_j w_{ij} \xi_j\right)=\xi_i, \quad \forall i. \label{stab}
\end{equation}
In this case, in fact, the rule \eqref{dynamics} will not produce any change. In order to achieve this result, we must change the value of the weights. It is easy to see that taking:
\begin{equation}
w_{ij}=\frac{1}{N} \xi_i\xi_j \label{weights}
\end{equation}
makes $\bm \xi$ a stable point. \\
We can now ask how to get the system to memorize many patterns, and, when asked to recall a pattern, recall the closest. Let $p$ be the total number of stored patterns, labelled by $\mu$. The simplest implementation consists in making the weights $w_{ij}$ a superposition of of the weights given in \eqref{weights}:
\begin{equation}
\label{Hebbs}
w_{ij}=\frac{1}{N}\sum_{\mu=1}^p \xi^\mu_i\xi^\mu_j, \qquad i\neq j.
\end{equation}
Since the diagonal weights $w_{ii}$ represent connections of neurons with themselves (see fig. \ref{net}), in order to avoid feedback effects, we assume $w_{ii}=0$, for all $i$.
\boxtext{This is called~\cite{Singer} the \textit{Hebb's rule}, because of the similarity with the observations made by the neuroscientist~\cite{Hebb}. Usually, instead of calculating all the weights at once from the patterns and assigning them to the network it is preferred to \textit{asynchronously update}, by selecting, at each time step, a random a unit $i$ to be updated and apply the rule. We will call \textit{Hopfield model} a binary valued associative memory using the Hebb rule and asynchronous updating~\cite{SantaFe}.}
The stability condition \eqref{stab} generalizes then to:
\begin{equation}
\sgn(x_i^\nu)=\xi_i^\nu, \quad \forall i,
\end{equation}
where $x_i^\nu$ is the net input to unit $i$ in pattern $\nu$. By applying Hebb's rule and extracting $\mu=\nu$ from the sum over $\mu$, we can apply $(\xi_j^\mu)^2=1$ and thus:
\begin{equation}
x_i^\nu= \sum_{j=1}^N w_{ij}\xi_j^\mu=\frac{1}{N} \sum_{j=1}^N\sum_{\mu=1}^p \xi_i^\mu\xi_j^\mu\xi_j^\nu=\xi _i^\nu + \frac{1}{N} \sum_{j=1}^N\sum_{\mu\neq \nu}^p \xi_i^\mu\xi_j^\mu\xi_j^\nu.
\end{equation}
We can immediately see that if the absolute value of the second term, called \textbf{crosstalk term}, is smaller than $1$ it cannot change the sign of $x_i^\nu$. If this condition holds for all $i$, we then have that the pattern $\bm \xi^\nu$ is a stable point of the network dynamics. Note that this condition is sufficient but not necessary, since a crosstalk term that has the same sign of $\xi_i^\nu$ will not change the sign of $x_i^\nu$, independently of its absolute value. 

\subsection{Storage Capacity} We consider the quantity $C_i^\nu$ given by minus the product of the crosstalk term with its associated pattern value $\xi_i^\nu$:
\begin{equation}\label{crosstalk}
C_i^\nu=-\xi_i^\nu \frac{1}{N} \sum_{j=1}^N\sum_{\mu\neq \nu}^p \xi_i^\mu\xi_j^\mu\xi_j^\nu.
\end{equation}
If $C_i^\nu$ is negative, it means that the crosstalk term has the same sign as the desired $\xi_i^\nu$ term, so, as observed before, it does no harm.
But if $C_i^\nu$ is positive and larger than 1, it changes the sign of $x^\nu_i$ and makes the pattern $\nu$ an unstable point for the $i$-th unit.
The quantity $C_i^\nu$ depends on the patterns, making it a random variable (in example \ref{ex5} we evaluate its associated probability distribution). We can therefore estimate the probability $P_e$ that the pattern $\nu$ is unstable for the $i$-th unit. 
\begin{equation}
P_e =\text{Prob}(C_{i}^\nu>1).
\end{equation}
Clearly $P_e$ increases as we increase the number $p$ of patterns that we try to store. We will see that fixing $P_{e}>0$ gives us a linear dependency between the number of storable patterns (within the choosen error probability) and the number of neurons.
For now, let us define the storage capacity of the associative memory as:
\begin{defi}[Storage capacity of a network]
The storage capacity of a network is the ratio $\alpha$ between the maximum number $p_{*}$ of patterns that can be stored without unacceptable errors, so that $P_e$ is less than a maximum error probability $P_{e_{*}}$, and the number of units $N$. 
\end{defi}
This definition of capacity extends the one used for the perceptron, where we assumed that every example have to be learned perfectly, so $P_e=0$.
\begin{ex}[Capacity of an associative memory] \label{ex5}
In this example we evaluate an upper bound to the storage capacity of a network of $N$ units storing $p$ random patterns, where each $\bm \xi^\mu$ is an array of $p$ independent values randomly chosen between the values $-1$ and $+1$ where each value has equal probability, $0.5$. \\
Let us start by showing that $C_{i}^\nu$ is a Gaussian centered in $0$ with variance $\sigma^2=p/N$.
\boxtext{We notice that $C_i^\nu$, as defined in \eqref{crosstalk} is the sum of the $Np$ independent and identically distributed random variables $\xi^\nu_i\xi_i^\mu\xi^\nu_j\xi_j^\mu$. Therefore, we can apply the \textbf{Central limit Theorem}:
\begin{equation}\label{Tlc}
Y=\lim_{M\to \infty}\left(\frac{1}{\sqrt{M}}\sum _{k=1}^{M}X_{k}-\mu\sqrt{M} \right) \, \Rightarrow \, Y \text{ has } N(0,1) \text{ distribution.}
\end{equation}
In this case we have $M= Np$ total independent variables $X_k^{i\nu}=- \xi^\nu_i\xi_i^\mu\xi^\nu_j\xi_j^\mu$, averaging over the indices $k=(j,\mu)$.
Moreover, the $X_k^{i\nu}$ variables have mean $\mu=0$ because the $\xi$'s are independent and each, taken alone, has mean $0$.\\
Therefore, in the limit of large $N$'s (and so, large $p$'s), the theorem applies yielding $C^\nu_i=\sigma Y^{i\nu}$, where $\sigma^2=p/N$ :
\begin{equation}\label{Tlc2}
C^\nu_i= \frac{1}{N}\sum_{\kappa=1}^{Np} X_k^{i\nu}=\sqrt{\frac{p}{N}} \frac{1}{\sqrt{Np}}\sum_{\kappa=1}^{Np} X_k^{i\nu}=\sigma Y^{i\nu}.
\end{equation}}
The probability $P_e$ that $C_{i}^\nu>1$ is thus:
\begin{equation}
P_e =\sqrt{\frac{N}{2\pi p}}\int_1^\infty e^{-Nx^2/2p}dx=\frac{1}{2}\left(1-\operatorname{erf}\left(\sqrt{N/2p}\right)\right).
\end{equation}
where the error function $\operatorname{erf}(x)$ is defined as follows:
\begin{equation}
\operatorname{erf}(x)=\frac{2}{\sqrt{\pi}}\int_0^x\exp(-t^2)\,dt.
\end{equation}
It is clear that for any fixed $P_e \neq 0$ it exists some value $\alpha>0$ such that:
\begin{equation}P_e=\frac{1}{2}\left(1-\operatorname{erf}\left(\sqrt{1/2\alpha}\right)\right), \quad \text{and} \quad p=\alpha N.\end{equation}
Then, for any finite error probability, also the maximum number of solvable problems $p_*$ is proportional to $N$. Here are some values of the proportionality constant $\alpha$ given the error probability $P_e$~\cite{SantaFe}:
\begin{equation}
\begin{array}{rr}
P_e & \alpha \\\hline
0.24 & 2 \\
0.16 & 1 \\
0.07 & 0.5\\
0.01 & 0.2\\
0.001&0.1\\
\end{array}
\end{equation}
It is important to note that with this calculation we can only establish an upper bound to the real capacity of the network. In fact with $P_e$ we are only measuring the \textbf{initial stability} of the patterns: choosing $P_e=0.01$ tells us that no more than $1\%$ of the \textbf{single bits} are initially unstable. But if about $1\%$ of the bits do flip, the new state may have more unstable bits and cause a positive feedback phenomenon which makes the whole memory unstable.
For example at $p = 0.138N$ only $0.37\%$ of the bits are initially unstable, though it turns out (through more sophisticated calculation) that about $1.6\%$ of the bits flip before a stable attractor is reached. \\
We want all the memorized patterns to be recalled perfectly. To do so it is worth studying the case $P_e\simeq 0$.
For example, for $P_e=0.01$ we have $p_*\simeq 0.2 N$, we can then expand $P_e$ for small values of $p/N$. The expansion yields:
\begin{equation}
\log(P_e) \simeq -\log (2\sqrt{\pi})-\frac{N}{2p}-\frac{1}{2}\log \frac{N}{2p}.
\end{equation}
Since each of the $p$ patterns contains $N$ bits, in order to get all of them right in the $99\%$ of the cases we need $(1-P_e)^{Np} > 0.99$, which is almost equivalent to $P_e<0.01/Np$. We then have
\begin{equation}
-\log (2\sqrt{\pi})-\frac{N}{2p}-\frac{1}{2}\log \frac{N}{2p}< \log 0.01 -\log Np.
\end{equation}
Which gives the $p_*=\frac{N}{4\log N}$, using $\log Np \simeq \log N$, so $\alpha =\frac{1}{4\log N}\xrightarrow{N\to \infty} 0$.\\
In summary, $p_*$ is proportional to $N$ if we are willing to accept a small percentage of errors in each pattern, but is proportional to $N/ \log N$ if we insist that all patterns be recalled
perfectly.
\end{ex}
\subsection{Statistical mechanics of a network}
In order to implement a statistical mechanical approach we need to associate an energy function to the dynamics, as Hopfield did in 1982 ~\cite{Hopfield}.
\begin{defi}[Energy function] The energy function $H$ is a function of the configurations $\{S_i\}$, with the fundamental property of decreasing or being constant as long as the system evolves according to its dynamical rule. For an associative memory, such an $H$ satisfies the property:
\begin{equation} \label{energy}
H=-\frac{1}{2}\sum_{i,j=1}^Nw_{ij}S_iS_j,
\end{equation}
where $w_{ii}=0$ for all $i$, as we have seen in eq. \eqref{Hebbs}.
\end{defi} 
First of all, let us prove that the dynamical rule (eq. \ref{dynamics}) can only decrease the energy.
\boxtext{
Each step of the dynamical evolution may or may not flip the value of a site $S_i$. Let us call $S_i'$ the evolved state of $S_i$, according to \eqref{dynamics}:
\begin{equation} S_i'=\sgn\left(\sum_j w_{ij} S_j\right) .
\end{equation}
If $S_i'=S_i$ the energy is unchanged, while in the other case $S_i'=-S_i$.
We can now assume, without loss of generality, that the flipped site is $S_1$.
Then the variation of energy associated with the flip of $S_1=-S_1'$ while the other $S_{j\neq 1}$ are left unchanged is:
\begin{equation}\label{endynamics}
\begin{aligned}
H'-H&=-\frac{1}{2}\sum_{ij}w_{ij}S'_iS'_j+\frac{1}{2}\sum_{ij}w_{ij}S_iS_j=\\
&=-\frac{1}{2}\sum_{ij}w_{ij}S'_iS_j+\frac{1}{2}\sum_{ij}w_{ij}S_iS_j=\\
&=\frac{1}{2}\sum_{j}S_j\sum_{i}w_{ij}(S_i-S_i')=\\
&=\frac{1}{2}\sum_{j}S_j\sum_{i\neq 1}w_{ij}(S_i-S_i')+\frac{1}{2}(S_1-S_1')\sum_{j}w_{1j}S_j=\\
& =0+ S_1 \sum_{j}w_{1j}S_j = S_1 S_1'=-S_1^2 = -1.
\end{aligned}
\end{equation}
}
We now assume that the weights are given by the Hebb's rule (equation \ref{Hebbs}). Equation \ref{energy} then becomes:
\begin{equation} \label{energy2}
H=-\frac{1}{2N}\sum_{\mu=1}^p\left(\sum_{i=1}^NS_i\xi^\mu_i\right)^2+\frac{p}{2}.
\end{equation}
Since eq. \eqref{Hebbs} does not naturally implements $w_{ii}=0$ we need to introduce a $\frac{p}{2}$ term to cancel out the diagonal terms. \\
Introducing a \textbf{temperature parameter} $T=\frac{1}{\beta}$, we can treat the $\{S_i\}_{i=1}^N$ as a thermodynamic ensemble. We then define the \textbf{partition function} of the system:
\begin{equation} \label{partition}
Z=\Tr_S\,\exp \big(-\beta H\big),
\end{equation}
where the trace $\Tr_S$ means the sum over all the possible states, $\{S_i=\pm 1\}_{i=1}^N$.
We again define $\alpha= p/N$ (memorized patterns over number of perceptrons in the net), and once again we will take the limit for large $N$. We will consider 2 cases $\alpha=0$, which means that $p$ is kept constant as $N\to \infty$, and $\alpha \neq 0$, which is when $p$ scales proportionally to $N$. 

\subsubsection{\texorpdfstring{Mean field theory for $\bm{\alpha=0}$}{Mean field theory for \textalpha=0}}
Let us define the \textbf{mean field parameters}:
\begin{equation} \label{mfpara}
m^\mu=\frac{1}{N}\sum_{i=1}^N \xi_i^\mu\langle S_i\rangle_\beta,
\end{equation}
where $\langle \hdots \rangle_\beta$ is the thermal average:
\begin{equation} \label{betaavg}
\langle A \rangle_\beta = \frac{1}{Z}\, \Tr_S[-\beta H A].
\end{equation}
The \textbf{free energy} per unit of the system (obtained through the calculations shown in Appendix \ref{h2}) is:
\begin{equation}\label{free} f(\beta, \bm m)=-\frac{1}{N\beta}\log Z = \frac{\alpha}{2} + \frac{1}{2}\bm m^2 -\frac{\log 2}{\beta}-\frac{1}{\beta N} \sum_{i=1}^N 	\log \cosh\big (\beta\bm m \cdot \bm \xi_i\big), \end{equation}
where $\alpha=p/N\to 0$ for large $N$. Minimizing $f$ with respect to $m^\mu$:
\begin{equation} \label{eq30} 0=  \frac{\partial f}{\partial m^\mu}(\beta, \bm m)=m^\mu -\frac{1}{N} \sum_{i=1}^N \xi_i^\mu \tanh \big (\beta\bm m \cdot \bm \xi_i\big), \end{equation}
yields the self-consistency equations for the mean field parameters:
\begin{equation}\label{mfeq} m^\mu =\frac{1}{N} \sum_{i=1}^N \xi_i^\mu \tanh \big (\beta\bm m \cdot \bm \xi_i\big). \end{equation}
We introduce now another average, $\langle \hdots \rangle_\xi$, this time over the distribution of patterns $\{\bm \xi_i\}_{i=1}^N $. Since the system is \textbf{self-averaging}, for any function of the input $A(\xi)$, we can deem the average over the input distribution equivalent to the average over units:
\begin{equation} \label{xiavg}
\langle A \rangle_\xi = \frac{1}{N}\sum_i A(\xi_i).
\end{equation}
\boxtext{The \textbf{self averaging} properties arise from the fact that as we go from neuron to neuron in the sum over index $i$ we are choosing $N$ independent $\bm \xi_i$ from the distribution $P(\bm \xi)$, which we assume is uniform over the $2^p$ possibilities. So for $N \gg 2^p$, which is guaranteed in the thermodynamic limit by $\alpha=0$, the average over sites is equivalent to an average over the distribution.}
The final form of the self-consistency equation is then:
\begin{equation} \label{selfcon} m^\mu =\Big \langle \xi_i^\mu \tanh \big (\beta\bm m \cdot \bm \xi_i\big) \Big\rangle_\xi. \end{equation}
We can now define the memory states in term of these $m^\mu$ parameters, solutions to the self-consistency equation \eqref{selfcon}. To choose which solutions to use, we make an \textit{ansatz} and take the equilibrium value $\langle S_i \rangle_\beta $ proportional to one of the stored patterns:
\begin{equation}\label{ans} \langle S_i\rangle_\beta =m \xi_i^\nu, \end{equation} 
as we have already seen in Section \ref{patmem} it gives stable solutions at $T=0$.
\begin{defi}[Memory states]
Those states that have a non-zero correlation with just one of the patterns, $\bm \xi^1$:
\begin{equation} \langle \xi^\mu \xi^\nu \rangle_\xi =\delta_{\nu,\mu}\delta_{\mu,1}.\end{equation}
 The vector $\bm$ for these states will then have the form:
\begin{equation} \label{memstates}
\bm m= (m, 0, \hdots, 0).
\end{equation}
\end{defi}
The self-consistency equation \eqref{selfcon} then reduces to (note that $\xi^1=\pm 1$ and $\tanh$ is odd):
\begin{equation} \label{selfcon2}
m^\mu =  \langle \xi^\mu \tanh \beta m \xi^1\rangle_\xi = \langle \xi^\mu \xi^1\rangle_\xi  \tanh \beta m = \delta_{\mu,1} \tanh \beta m .
\end{equation}
This equation has non-zero solutions $\forall T < T_c=1$ that imply stable memory states (see Appendix \ref{h3}). There exist also more complicated solutions to the mean field equations but it can be proven that the corresponding \textbf{spurious states} are all unstable for $T>0.46$.

\subsubsection{\texorpdfstring{Mean field theory for $\bm{\alpha\neq 0}$}{Mean field theory for nonzero \textalpha}}
As we saw in example \ref{ex5} if $p$ is the order of $N$ (so $p = \alpha N$) we have to expect $P_e\neq 0$. In order to make this theory mathematically tractable we have to make some assumptions: the most important one is that, even if the number of stored patterns increases with $N$, we keep \textbf{finite} and equal to $s$ the number of stored patterns -  called \textbf{condensed patterns} - that have non-zero correlation with the memory states. The order parameter $\bm m$ at the saddle point takes now the form $\bm m= (m^1,\hdots, m^s ,0,\hdots,0)$.
Other assumptions, together with the rest of the very cumbersome calculations for the case $\alpha \neq 0$, are detailed in Appendix \ref{h4}.
This time, alas, the \textbf{order parameters} $m^\mu$ are not enough to describe the system, we should instead add the parameters $q$ and $r$ (their rationale behind these parameters is explained in Appendix \ref{h4b}):
\begin{equation} \label{mfeq2}
m^\mu=\frac{1}{N}\sum_{i=1}^N \xi_i^\mu\langle S_i\rangle_\beta, \qquad q=\left \langle \frac{1}{N}\sum_{i=1}^N \langle S_i\rangle_\beta^2 \right \rangle_{\!\!\xi}, \qquad r=\frac{1}{\alpha}\sum_{\mu > s} \left\langle (m^\mu)^2\right\rangle_\xi.
\end{equation}
We will also need a Gaussian random variable $z$ to represent the effects of the uncondensed patterns $\mu>s$.\\
As in the $\alpha=0$ case, from the partition function $Z$ we can write a free energy per unit function:
\begin{equation}\label{fepu}
\begin{aligned}F(\bm m, \beta, q,r,z)/N&=-\frac{1}{\beta N}\langle \log Z \rangle_\xi=\\ &= \frac{1}{2}\alpha+ \frac{1}{2}\bm m^2 + \frac{\alpha}{2\beta}\left(\log[1-\beta(1-q)-\frac{\beta q}{1-\beta(1-q)}\right)+\\
&+\frac{1}{2}\alpha \beta r(1-q)-\frac{1}{\beta} \log 2 -\frac{1}{\beta}\left\langle\log \cosh \beta(z\sqrt{\alpha r}+\bm m \cdot \bm \xi) \right \rangle_{\xi,z}.
\end{aligned}
\end{equation}
By looking for maxima and minima of $F/N$  we find the self-consistency equations for the parameters $m^\mu,r,q$:
\begin{equation} \label{sys} \begin{cases}
m^\mu=\langle \xi^\mu \tanh\beta(z\sqrt{\alpha r}+\bm m \cdot \bm \xi)  \rangle_{\xi,z}\, ,\\
q=\langle \tanh^2 \beta(z\sqrt{\alpha r}+\bm m \cdot \bm \xi)  \rangle_{\xi,z}\,,\\
r=\frac{q}{(1-\beta(1-q))^2}.
\end{cases} \end{equation}
Solving these equations for pure memory states $\bm m =(m,0, \hdots,0)$, we find that, varying temperature, the $m$ parameter incurs a double phase transition. More details in Appendix \ref{h5}.
\newpage
\section{Gardner's theory} 
\begin{figure}[h!]
\centering
\includegraphics[width=0.45\linewidth]{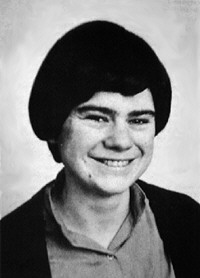}
\vspace{0.4cm}
\caption{Elizabeth Gardner (25 August 1957 – 18 June 1988) was a British theoretical physicist, best known for her groundbreaking work on spin glasses, phase transitions and disordered networks.\cite{Enderby}}
\label{EG}
\end{figure}

This theory\cite{Gardner}, developed by E. Gardner's \ref{EG} starting from the Hopfield approach, aims to evaluate the capacity as the fraction of volume of the weights that implement the desired solution. But instead of using the $S_i$ as statistical-mechanical variables, it uses the weights $w_{ij}$.\\
We will assume continuous weights that satisfy the spherical normalization:
\begin{equation}\label{norm} \sum_{j=1}^Nw_{ij}^2=N, \quad \forall i =1, \hdots, N. \end{equation}

\subsection{Capacity of a simple perceptron}
\begin{figure}[h]
\centering
\includegraphics[width=0.6\linewidth]{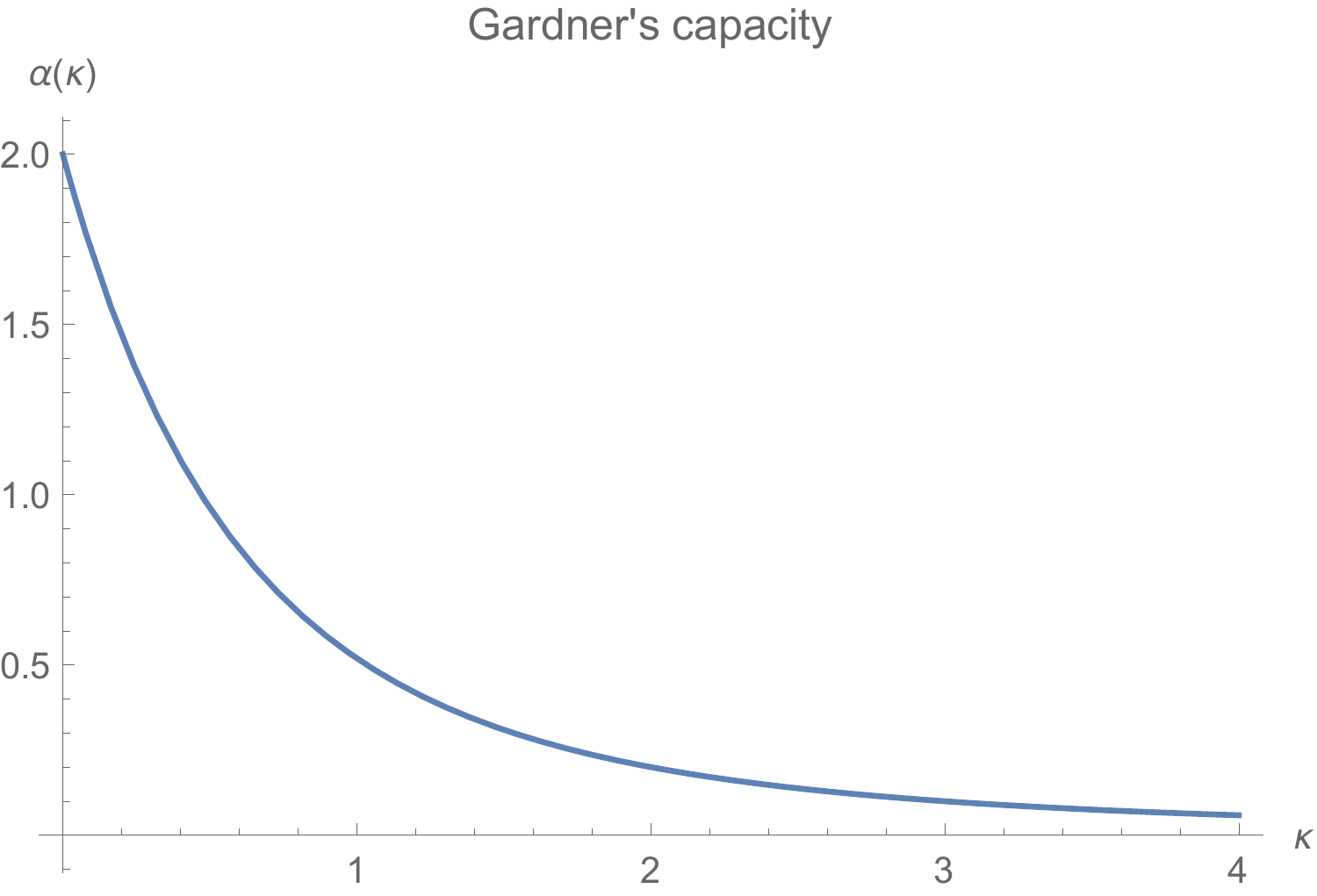}
\caption{Trend of the simple perceptron capacity \textit{à la Gardner}, according to equation \eqref{alpha}. }
\label{alfa}
\end{figure}

Let us start by considering a simple perceptron $i$ with $N$ binary inputs $\xi_j$, $j=1,\hdots, N$  that outputs $\zeta_i$, defined as:
\begin{equation}\zeta_i= \sgn\left(N^{-1/2}\sum_{j=1}^N w_{ij} \xi_j\right). \end{equation}
We want to find the volume of the $w_{ij}$ such that this equation is satisfied for each one of a set of $p$ examples $\bm \xi^\mu \to \zeta_i^\mu$, this condition is then equivalent to the following inequality:
\begin{equation}\label{nonsolid}\zeta_i N^{-1/2}\sum_{j=1}^N w_{ij} \xi_j >0. \end{equation}
To make the perceptron solid enough to correct small errors in the input patterns, we can make this condition more restrictive, for example by introducing a \textbf{margin size} $\kappa>0$:
\begin{equation}\label{solid} \zeta_i N^{-1/2}\sum_{j=1}^N w_{ij} \xi_j >\kappa. \end{equation}
In fact, for $\kappa$ large enough, if \eqref{solid} is satisfied, \eqref{nonsolid} will be satisfied even if some bit of the input are given wrong.\\
The ratio, between the volume of the weights that satisfy both the normalization \eqref{norm} and solve the $p$ examples \eqref{solid} and the volume of weights that satisfy just the normalization, is then:
\begin{equation}\label{volume} V=\frac{\int\prod_{ij}dw_{ij} \, \prod_{i\mu}\Theta(\zeta_i^\mu N^{-1/2}\sum_{j=1}^N w_{ij} \xi_j-\kappa)\, \prod_i \delta(\sum_jw_{ij}^2-N)}{\int\prod_{ij}dw_{ij} \, \prod_i \delta(\sum_jw_{ij}^2-N)}\,. \end{equation}
This expression is similar to a partition function $Z$, where the exponential distribution is replaced by an all-or-nothing step function. In this case we introduce the order parameters $q,F,E$, where $q$ is the \textbf{correlation between the different weights solutions}
and in Appendix \ref{h6} we find we can express $E$ and $F$ in terms of $q$.
Given different perceptrons $\rho$ and $\sigma$ implementing the same problem, we can represent $q$ as:
\begin{equation}
 q=\frac{1}{N}\sum_{j=1}^Nw_j^\rho w_j^\sigma, \quad \forall \rho \neq \sigma.
\end{equation}
The quantity $G=\frac{1}{N}\langle \log V\rangle_{\xi, \zeta}$ will then give us, through partial derivatives, self-consistent equations that are solved by those configurations of weights that solve the problem. 
For a simple perceptron, the $G$ function reads:
\begin{equation}\label{gfunc} G(\alpha, \kappa, q) =\alpha \int\frac{dt}{\sqrt{2\pi}}e^{-t^2/2}\log \left[\int_{\frac{t\sqrt{q}+\kappa}{\sqrt{1-q}}}\frac{dz}{\sqrt{2\pi}}e^{-z^2/2}\right]+\frac{1}{2}\log 2\pi+\frac{1}{2}\log (1-q)+\frac{q}{2(1-q)}+\frac{1}{2}\,. \end{equation}
If the capacity $\alpha=p/N$ is small, there is a large region of $w$ space that solves the problem, then different solutions are usually uncorrelated and therefore $q\sim 0$. As we increase $\alpha$, there will be less and less weight values that solve all the problem, so the overlap between this solutions will increase. Finally, for the optimal perceptron (the one with the largest capacity) $q$ will become $1$.   
We should then calculate:
\begin{equation}\frac{\partial G}{\partial q}(\alpha, \kappa, q)\bigg\vert_{q\to 1}=0.\end{equation}
The limit gives us the expression of the optimal $\alpha$ given the margin size $\kappa$:
\begin{equation}\label{alpha}\alpha(\kappa)=\left[\int_{-\kappa}^\infty \frac{dt}{\sqrt{2\pi}}e^{-t^2/2}(t+\kappa)^2\right]^{-1}.\end{equation}
As can be seen in figure \ref{alfa}, this expression has its maximum for $\kappa=0$ at $\alpha=2$, in agreement with the results of Section \ref{percap}.

\subsection{Capacity of a \textit{quadratic} perceptron}
Following the work of Lewenstein et al. we now apply the same calculations to a perceptron where the output is given by:
\begin{equation}\zeta_i= \Theta\left(N^{-1}\left|\sum_{j=1}^N w_{ij} \xi_j\right|^2 - \kappa \right), \end{equation}
where weights lay on a sphere with the following normalization:
\begin{equation}\label{norm2} \sum_{j=1}^N\left\vert w_{ij}\right\vert^2=N, \quad \forall i =1, \hdots, N. \end{equation}
As we understood in example \ref{ex3}, as long as $\kappa>0$, the result does not depends on $\kappa$. We can therefore treat it in the same way we treated the margin size in \eqref{solid}: as a convergence parameter. Moreover since $ \zeta_i=0,1$, we can already drop it.
\begin{equation}\label{qperc}N^{-1}\left|\sum_{j=1}^N w_{ij} \xi_j\right|^2 > \kappa, \end{equation}
and once again we evaluate the ratio between the volume of the weights that satisfy both the normalization \eqref{norm2} and solve the $p$ examples \eqref{qperc} and the volume of weights that satisfy just the normalization \eqref{norm2}:
\begin{equation}\label{qsolid} V=\frac{\int\prod_{ij}dw_{ij} \, \prod_{i\mu}\Theta\left(N^{-1}\left\vert \sum_{j=1}^N w_{ij} \xi_j\right\vert^2 -\kappa\right)\, \prod_i \delta(\sum_j|w_{ij}|^2-N)}{\int\prod_{ij}dw_{ij} \, \prod_i \delta(\sum_j|w_{ij}|^2-N)}\,. \end{equation}
Our aim now is to evaluate the \textit{free energy function} $G$, in a similar fashion to what we did for the simple perceptron, \eqref{gfunc}.
We thus need to evaluate $G=\frac{1}{N}\langle \log V \rangle_{\xi}$, but, in order to do so, we shall assume a distribution for the inputs $\xi$.
Since \eqref{qsolid} is the partition function of a classical spin glass \cite{Lewenstein}, where the weights $w$ are classical spin variables and $\langle\hdots\rangle_{\xi}$ is a \textit{disorder average}, it is natural to assume an Ising distribution of the $\xi$'s.
Following the calculations detailed in Appendix \ref{h7}, we find the expression for the $G$ function, which is very similar to the one of the previous case, :
\begin{equation}\label{qg} G(\alpha, \kappa, q) =\alpha \int\frac{dt}{\sqrt{2\pi}}e^{-t^2/2}\log \left[\int_{\frac{t{q}+\kappa}{\sqrt{1-q}}}\frac{dz}{\sqrt{2\pi}}e^{-z^2/2}\right]+\frac{1}{2}\log 2\pi+\frac{1}{2}\log (1-q)+\frac{q}{2(1-q)}+\frac{1}{2}\,. \end{equation}
There is just a slight change in the behaviour of the \textit{correlation parameter} $q$ in the Gaussian integral.
We then differentiate $G$ and take the limit $q\to 1$ (the limit of \textit{dense solutions}), finding the expression of $\alpha$ as a function of $\kappa$. This expression is exactly like the one for the simple perceptron, eq. \eqref{alpha}, except for a factor $2$. It thus reaches a maximum value of $\alpha=4$ for $\kappa=0$.\\
From \ref{ex3} and \ref{ex4} we expected the \textit{quadratic} perceptron to have an higher capacity then the simple one; as it turns out, it is twice as much.
\newpage
\section{Conclusions}
The thesis focused upon the capacity of simple perceptrons in the classical and quantum regimes. In order to do it, we used the Gardner's statistical approach, which can be summarized as follows:
\begin{enumerate}
    \item In order to obtain the \textit{free energy} of the system, evaluate the n-replica partition function.
    \item Evaluate the average, assuming the desired distribution.
    \item Assume the replica symmetry. Note: this assumption may fail in some cases \cite{Lewenstein}.
    \item From the free energy we can easily obtain the equations for the model quantities.
    \item If needed, we can make other assumptions to get rid of variables that are not useful, like taking the limit of dense solutions $q\to 1$.
\end{enumerate}
The results obtained with this technique are often found to be in good accordance with the results observed in simulations \cite{SantaFe} or through other means (for example, the geometric calculation of the single perceptron capacity).\\
Thanks to the powerful Gardner's approach, in this thesis we have been able to investigate two different perceptron models, comparing their pros and cons:
The simple perceptron, despite having a lower capacity, is easy to implement in classical computers and is very versatile, being able to overcome its limitations when connected to other perceptrons to obtain more powerful \ref{ffnet} tools. \\
The \textit{quadratic} perceptron, on the other hand has double the capacity of the single one and, beside being implementable on classical hardware, can also be implemented \textit{straightforwardly} on quantum processors, even if without obvious advantages.
Anyway, we can now answer the question "how inherently quantum is the perceptron introduced in ~\cite{Macchiavello}". \\
The authors propose a quantum algorithm to implement a specific mathematical model, the \textit{quadratic} perceptron, which can also be implemented with classical computers, and even though the algorithm developed is a full-fledged quantum algorithm - for example, the hypergraph states provided by the circuit are often entangled - it is not advantageous in terms of complexity compared to a classical algorithm for the same \textit{quadratic} perceptron, as we showed in \ref{comp2} and \ref{comp3}).\\
So not only the same \textit{quadratic} perceptron model can be implemented on classical hardware, but in his quantum implementation  fails to deliver the complexity reduction which characterizes quantum algorithms like Shor's \cite{Shor} and Grover's \cite{Grover}.\\
As for the mathematical model of the \textit{quadratic} perceptron itself, it has nothing inherently quantum, but instead, the model gives rise to a question. As we found from Gardner's theory, the limit $\kappa \to 0$ is fundamental to obtain the maximum capacity from the \textit{quadratic} perceptron, but unfortunately, in some cases, like in \ref{ex5}, the limit can not be taken.\\
In fact, in the limit $\kappa \to 0$, the perceptron isolates orthogonal states from all the other, thus creating two subspaces with different dimension, making the model possibly more suited for storing values than for classification purposes.
Anyway, whether this model can be considered a \textit{strictu senso} classifier lies outside the scope of this thesis and would require further research.

\newpage
 \appendix 
 \section{Number of problems that a perceptron can solve}
 \label{h1}
 $C(p, N)$ can be calculated using a recurrence relation, 
 \begin{itemize}
\item Let us assume that we have already arranged $p-1$ points and that we know all the $C(p-1,N)$ different ways of putting a hyperplane between them. Please note that two hyperplanes are considered distinct if they yield a different shattering of the patterns. 
\item We add a $p$-th point. The $C(p-1,N)$ known hyperplanes can be divided into two categories:
\begin{enumerate}
\item Those that can be made pass through the $p$-th point without altering the separation between the previous points.
\item Those which, on the other hand, cannot pass through the new point without giving rise to a different separation between the previous $p-1$ points.
\end{enumerate}
We call the number of these hyperplanes $r_1$ and $r_2$ respectively. We observe that:
\begin{equation} r_1 + r_2 = C (p-1, N).\end{equation}
\item The $r_2$ hyperplanes that cannot be made pass through the $p$-th point, after adding the latter, give the same number of separations that they gave before.
\item Instead, the $r_1$ hyperplanes that can pass through the new point will each give rise to $2$ different separations, as they can be slightly rotated away from the $p$-th point leaving it either in the lower or upper half-space associated with the hyperplane. Therefore: 
\begin{equation}C(p,N)= 2r_1 + r_2 = C (p-1, N) + r_1.\end{equation}
\item r1 is the number of hyperplanes separating $p-1$ points with the additional requirement that the hyperplane passes through a given external point, but this constraint is equivalent to having a space with one dimension less, so: 
\begin{equation}r_1 =C (p-1, N-1).\end{equation}
\item We found the recurrence relation:
\begin{equation} C(p, N) = C (p-1, N) + C (p-1, N-1).\end{equation}
It can be easily checked that this relation is satisfied by
\begin{equation} C(p, N) =2\sum^{N-1}_{k=0}\binom{p-1}{k},\end{equation}
where we assume $\binom{p-1}{k}=0$  for $k \geq p$.
\end{itemize}

\newpage
\section{Macchiavello's quantum perceptron}
\label{h0}
This quantum information-based algorithm for the implementation of the \textit{quadratic} perceptron is based on an original procedure to generate multipartite entangled states, called \textit{hypergraph states generation subroutine}, or HSGS. \\
The gates used in the subroutine, are listed below:
\begin{itemize}
    \item $X$, the NOT gate: it changes the standard basis states vectors $|0\rangle$ and $\vert1\rangle$ of a qubit one into the other. It is implemented by the unitary Pauli matrix $\sigma_1$.
    \item $Z$, the bit-flip gate: it switches the sign of the $|1\rangle$ component of the state of the qubit on  which is applied.
    It is implemented by the Pauli matrix $\sigma_3$.
    \item $H$, the Hadamard gate: sends $|0\rangle$ into $\frac{|0\rangle+|1\rangle}{\sqrt{2}}$ and $|1\rangle$ into $\frac{|0\rangle-|1\rangle}{\sqrt{2}}$.\\
    When applied to the $q$ qubit state $|0\rangle^{\otimes q}$ it yields the uniform superposition $2^{-q/2}\sum_{j=0}^{2^q-1}|j\rangle$.
    \item $CU$ the controlled gate: it acts on two qubits, the first being used as a control qubit and the other one as a target qubit. It applies the unitary operator  $U$ to the target qubit only when the control qubit is $|1\rangle$: $CU|0,\psi\rangle=|0,\psi\rangle$, but $CU|1,\psi\rangle=|1,U\psi\rangle$. \\
    It is implemented by the $4\times4$ matrix: $$CU=
\vert 0\rangle\langle 0\vert\otimes I+\vert 1\rangle\langle 1\vert\otimes U=\begin{pmatrix}I&0\cr0&U\end{pmatrix}\ .
$$
    \item $C^kU$ the multi-controlled gate: it consists of $k+1$ qubits, the last qubit is the target qubit while the first $k$ ones act as control qubits. It applies, it applies $U$ to the target qubit only when all the control qubits are in the $|1\rangle$ state, i.e. the $|1\rangle^{\otimes q}$ state, which is written as $|2^q-1\rangle$ in decimal notation. 
   
\end{itemize}

\subsection{HSGS implementation}
The HSGS subroutine represents binary valued vectors, such as binary inputs $\bm \xi$ and binary weights $\bm w$, as the coefficients of a quantum state, which can be the output of a quantum circuit consisting of a suitable array of unitary gates. The quantum state of interest in the following are called \textit{hypergraph states}.

\begin{defi}[Hypergraph states \cite{Macchiavello2}]
Given a binary Boolean function $f:\{0,1\}^q \to \{0,1\}$, we define an hypergraph state as the following linear superposition::
\begin{equation}
|f\rangle=2^{-q/2}\sum_{j=0}^{2^q-1}(-)^{f(j)}|j\rangle.
\end{equation}
For any fixed $N=2^q$, there exist $2^{N}=2^{2^q}$ possible Boolean functions, that correspond to as many independent states.
\end{defi}

The hypergraph states play a very important role in quantum computing. For example, in Grover's search algorithm \cite{Grover}, $f$ is the marker function that identifies the object to find. They are also used in the Deutsch-Jozsa algorithm \cite{DJ}, whose purpose is precisely to identify whether a Boolean function $f$ is constant or \textit{balanced} (equal numbers of $0$ and $1$'s).

\begin{defi}[HSGS]
We will now describe the procedure that implements any Boolean function, and therefore any binary valued vector, as an hypergraph state. The result of this procedure will be a set of $C^kZ$ quantum gates that applied to the uniform superposition $2^{-q/2}\sum_{j=0}^{2^q-1}|j\rangle$ state, give the desired hypergraph state. Conceptually, the procedure start from the hypergraph state and at each iteration gets rid of some of the $-$ sign.
\end{defi}

\begin{enumerate}
\item We start with the desired hypergraph state, $|f\rangle=2^{-q/2}\sum_{j=0}^{2^q-1}(-)^{f(j)}|j\rangle$, 
we consider the components $|j\rangle$ such that $f(j)=1$ and the binary representation of $j$ presents only a single $1$ digit (the other digits being $0$). Let us assume there are $l$ such components, then $\{k_1, \hdots, k_l\}$ are indices that mark the position of that $1$ digit in each of the $l$ components.
\item We then apply $l$ gates $Z$ to $|f\rangle$. We apply those gates to the qubit indexed by $(k_1, \hdots, k_l)$, so, with a convenient notation, we apply the $\{Z_{k_1}, \hdots,Z_{k_l}\}$ gates to the circuit. This action will flip the sign of all the components $|j\rangle$ that have a digit $1$ in one of the positions $(k_1, \hdots, k_l)$. If a component $|j\rangle$ has a $1$ in two of the positions $(k_1, \hdots, k_l)$, than one sign will be flipped twice, and so on.
\item We now repeat step 1, but instead of considering the components such that $j$ only have single $1$ digit in binary notation, we consider those with two digits $1$ (still $f(j)=1$, of course). We then have a set of couples of indices. For each of these couples $(k_{l1}, k_{l2})$ we apply the corresponding two-qubit gate $CZ$ on the $k_{l1}$ and $k_{l2}$ qubits. 
Since these gates commutes and are symmetric with the respect to target and control, the order of the couples and even the one of the indices inside a couple does not matter. We then have another list of gates, $\{CZ_{k_{l1},k_{l2}}\}_{(k_{l1}, k_{l2})}$, that we applied to $|f\rangle$.
\item We repeat these procedure with three-qubit gates $C^2Z$ (once again, invariant under permutations of target and controls) and so on, until $C^{q-1}Z$, where $q$ is the number of qubit. 
\item This way, we get rid of the $-$ sign in the components with less digit $1$ first and, even if applying a $C^kZ$ can create $-$ signs in previously $+$ signed components, it can do so only on states with $k+1$ digits $1$. So we always end up with the uniform superposition $2^{-q/2}\sum_{j=0}^{2^q-1}|j\rangle$ and the $q$ lists of $C^kZ$ gates needed to implement $|f\rangle$.
\end{enumerate}
Let us call $U_f$ the quantum gate that implement $|f\rangle$ from $|0\rangle^{\otimes q}$, that is the gate formed by first applying an $H$ gate to each qubit and then the listed $C^kZ$ gates.

\subsection{Functioning}
\begin{figure}[h!]
\vspace{0.5cm}
\centering
\includegraphics[width=0.75\linewidth]{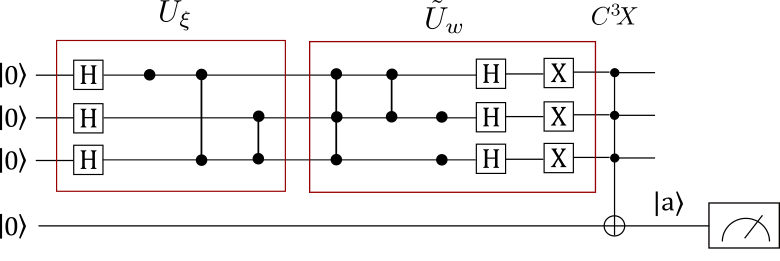}
\caption{The quantum perceptron that implement the scalar product between \\$\bm \xi =(1,-1,1,-1,1,1,-1,-1)$ and $\bm w =(1,1,-1,-1,1,-1,1,1)$}
\label{Macchiavello2}
\end{figure}

We now have all the tools we need to implement the $N$-input \textit{quadratic} perceptron (see fig. \ref{Macchiavello2}). The inputs are binary ($\xi=\pm 1$).
\begin{enumerate}
\item We create a quantum circuit with $q+1$ qubit, where $q=\lceil\log_2 N\rceil$, initializing all of them in the $|0\rangle$ state.
\item We then apply the gate $U_\xi$ that implements the desired inputs on the first $q$ qubits, making the state of the circuit:
\begin{equation}
|\xi\rangle = U_{\xi}\,|0\rangle^{\otimes q} = \frac{1}{\sqrt{N}}\sum_{j=0}^{N-1} \xi_j|j\rangle.
\end{equation}
\item We now consider the gate $\tilde U_w$ made by a $U_w^\dagger$ gate - defined as before, with $w_j$ instead of $\xi_j$ as coefficients - and $q$ $X$ gates, one for each of the qubits, such that: 
\begin{equation}
|w\rangle= \frac{1}{\sqrt{N}}\sum_{j=0}^{N-1} w_j|j\rangle= \tilde U_w^\dagger \, |1\rangle ^{\otimes q} .\end{equation}
We then apply this $\tilde U_w$ gate to the circuit and call the resulting state: 
\begin{equation}
|\psi\rangle=\tilde U_w\, |\xi\rangle =\tilde U_w \, U_\xi \,| 0\rangle^{\otimes q}=\sum_{j=0}^{N-1}c_j\, |j\rangle,
\end{equation}
for some coefficients $c_j$, with $j=0, \hdots, N-1$.
\item This way, we are taking the scalar product between the two vectors in the coefficient $c_{N-1}$ of the $|1\rangle ^{\otimes q}=|N-1\rangle$ component of the circuit's state. In fact the last component $c_{N-1}$ is :
\begin{equation}
c_{N-1}=\langle N-1 |\psi \rangle=\langle N-1|\tilde U_w\, \tilde U_w^\dagger |\psi \rangle=\langle w \, | \,\xi \rangle = \frac{1}{N}\,\bm w\cdot \bm \xi.
\end{equation} where we used the unitarity of quantum gates $\tilde U_w\, \tilde U_w^\dagger=I$.
\item To collect this result, we apply a $C^qX$ to the circuit, using the $q$ qubits in the $|\psi\rangle$ state as controls and the ancilla as target. As the state $|a\rangle =|0\rangle$ of the ancilla will be flipped to $|1\rangle$ only when all the controls are $|1\rangle$, that is, only for the $|N-1 \rangle$ component. The state of the network is then:
\begin{equation}
|\psi, a\rangle = \sum_{j=0}^{N-2}c_j|j,0\rangle +c_{N-1}|N-1, 1\rangle.
\end{equation}
\item If we now measure the ancilla state, we will find it equal to $|1\rangle$ with probability:
\begin{equation}
P_1=|c_{N-1}|^2=  \frac{1}{N^2}\,|\bm w\cdot \bm \xi|^2.
\end{equation}
We will measure $|0\rangle$ the rest of the time, $P_0=1-P_1$.
\end{enumerate}

\subsection{Complexity}\label{comp3}
To estimate the complexity of the circuit - namely, the number of elementary one and two-qubit gates needed to implement it - we evaluate the complexity of the $U_\xi$ operator.
In the case of $q$ qubit, and therefore $N = 2q$ input, we may have to apply up to $\binom{q}{k}$ gates $C^kZ$, for each $k$ ranging from $0$ to $q-1$.
The complexity of a gate $C^kZ$ in a quantum circuit depends on the basis of elementary gates chosen, anyway it can be approximated by $O(k^2)$ \cite{QGDec}.
Overall, therefore, the circuit has complexity:
\begin{equation}(q-1)!\sum_{k=0}^{q-1}\frac{k^2}{k!(q-1-k)!} \simeq q\, 2^q =N\log N.\end{equation}

\newpage
\section{\texorpdfstring{Mean field theory for $\bm{\alpha=0}$}{Mean field theory for \textalpha=0}}
\label{h2}
In order to implement the mean field, we add $p$ auxiliary external fields $h^\mu$ to the Hamiltonian. These field will be set to $0$ later, after they have served their purpose. The partition function then becomes:
\begin{equation} \label{partition2}
Z=e^{-\beta p/2} \Tr_S\,\exp \left(\frac{\beta}{2N} \sum_{\mu=1}^p\left(\sum_{i=1}^NS_i\xi^\mu_i\right)^{\!\! 2} +\beta \sum_{\mu=1}^p h^\mu \sum_{i=1}^NS_i\xi^\mu_i  \right),
\end{equation}
The argument of the exponential is quadratic, which means we can not simply factorize the trace over the system into $N$ independent traces $\Tr_{S_i}$ over the single perceptron. To linearize the exponential argument, we can apply the \textit{Gaussian integral trick}:
\begin{equation} \label{gauss}
\int_{-\infty}^{+\infty}dx\,e^{-ax^2\pm b x}=\sqrt{\frac{\pi}{a}}e^{b^2/4a}, \qquad \text{with }a>0,
\end{equation}
at the expense of introducing $p$ auxiliary variables $m^\mu$ and just as many integrals over these.
After some mathematical manipulation we can thus write $Z$ as:
\begin{equation} \label{partition3}
Z=\left(\frac{N\beta}{2\pi}\right)^{p/2}\int d^p\!m \, e^{-\beta N f(\beta, \bm m)},
\end{equation}
where:
\begin{equation} \label{freeenergy1}
f(\beta, \bm m)=\frac{p}{2N}+\frac{\bm m^2}{2} - \frac{1}{\beta N}\sum_i \log (2\cosh ( \beta(\bm m +\bm h)\cdot \bm \xi_i)).
\end{equation}
Let us approximate the integral with the \textit{saddle point method}: for $N\to \infty$, the integral is dominated by those values $m_0$ where $f$ is small, so:
\begin{equation} \label{saddlepoint1}
\int dm e^{-Nf(m)}  \simeq \int_{-\infty}^{+\infty} e^{-N\left (f(m_0) + 0 + \frac{1}{2} f''(m_0)(m-m_0)^2\right)} = e^{-Nf(m_0)} \sqrt{\frac{2\pi}{Nf''(m_0)}}
\end{equation}
Moreover, for large $N$ the result is dominated by the exponential, so:
\begin{equation} \label{partition4}
Z=\left(\frac{N\beta}{2\pi}\right)^{p/2}  e^{-\beta N f(\beta, \bm m)}.
\end{equation}
Here we renamed $\bm m$ the saddle points values $\bm m_0$. Since $m$ are saddle points, we have our mean field equation \eqref{eq30}:
\begin{equation} \label{mf}
0= \frac{\partial f}{\partial m^\mu} = m^\mu - \frac{1}{N}\sum_{i=1}^N \xi_i^\mu \tanh (\beta(\bm m +\bm h ) \cdot \bm \xi_i).
\end{equation}
 \subsection{Stable memory states}
  \label{h3}
We have observed that under the \textit{ansatz} \ref{ans} and in the limit of $p\ll N$, the self-consistency equation \ref{selfcon} takes the form \eqref{mfeq}.
In order to prove that this equation has stable solutions for $\beta \geq 1$ ($T\leq T_c=1$) we analyze the free energy function for pure states $\bm m = (m, 0, \hdots, 0)$. Plugging it into the definition of free energy \eqref{free} gives (neglecting constant terms):
 \begin{equation}
f(\beta, m)=\frac{1}{2}m^2- \frac{1}{\beta} \log \cosh \beta m.
\end{equation}
For $\beta \geq 1$, this function shows a double well. Solutions around the minima $m\simeq \pm 1$ are therefore stable.
Instead, for $\beta \leq 1$  the function shows a single well at $m=0$, so stable solutions will have $m= 0$.
The phase transition of the system can therefore be studied in the context of \textbf{Landau theory}.
\begin{figure}[!h]
\vspace{0.5cm}
     \centering
     \begin{subfigure}[l]{0.4\textwidth}
       \centering
	\includegraphics[width=\textwidth]{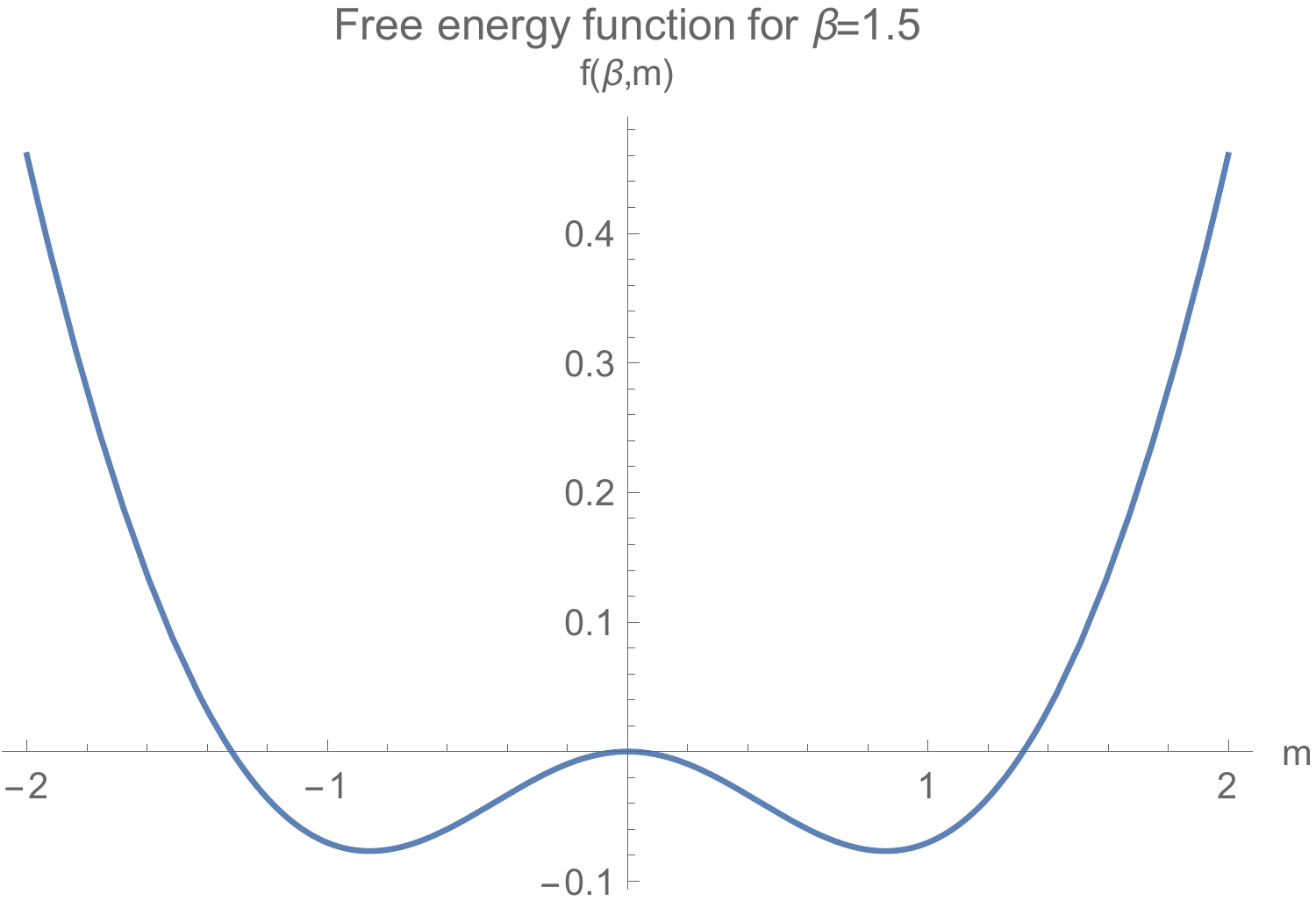}
		\captionsetup{width=\textwidth}	
	\caption{Free energy function for pure states when $\beta >1$.}
	\label{ffig}
     \end{subfigure}
     \quad
     \begin{subfigure}[r]{0.4\textwidth}
	\centering
	\captionsetup{width=\textwidth}	
	\includegraphics[width=\textwidth]{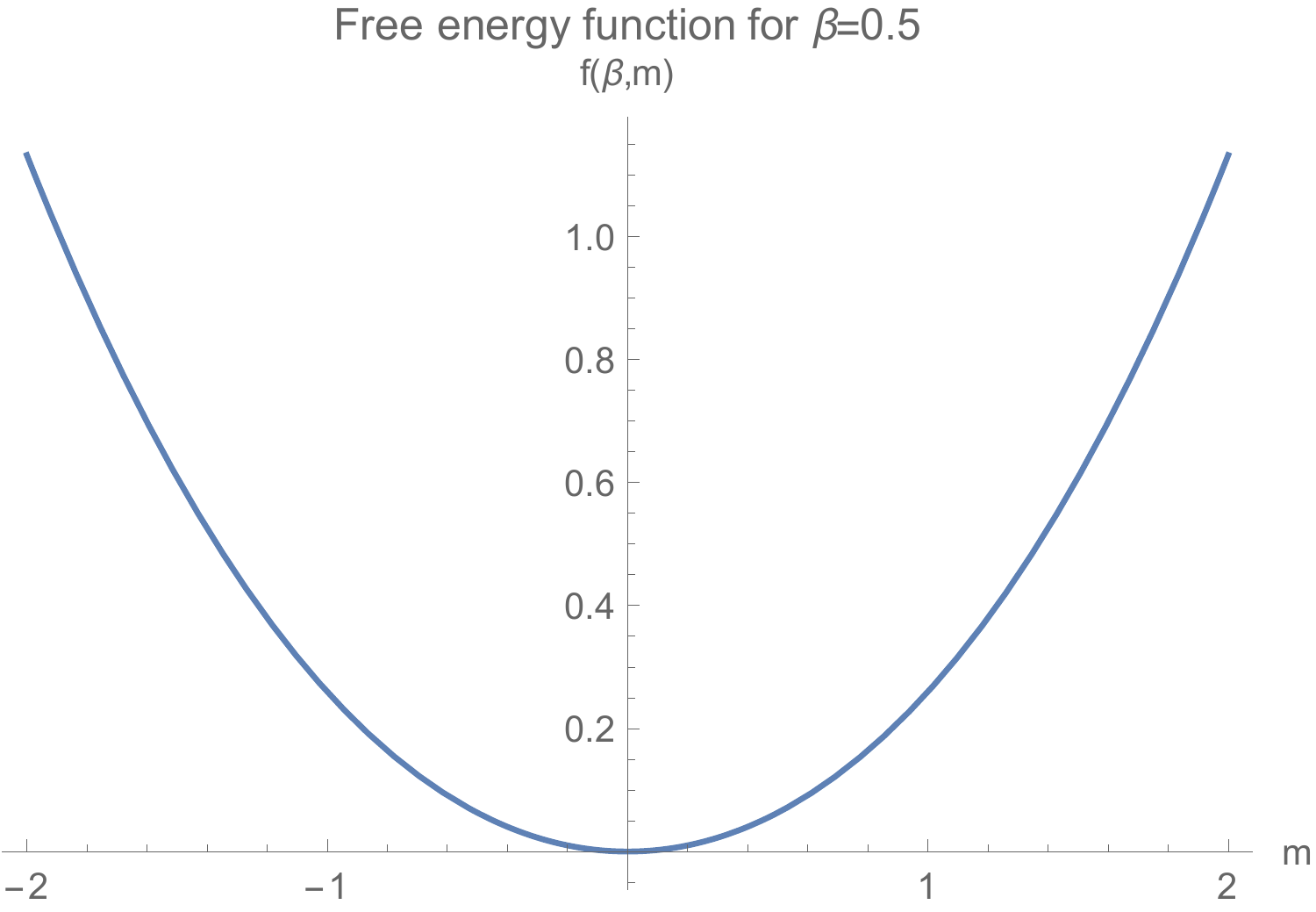}
	\caption{Free energy function for pure states when $\beta < 1$.}
	\label{phase}
     \end{subfigure}
\end{figure}
 \newpage 
\section{\texorpdfstring{Mean field theory for $\bm{\alpha\neq 0}$}{Mean field theory for nonzero \textalpha}}\label{h4}
  In order to write the free energy function, we need to calculate $\log Z$. Since now $Z$ depends on the set of pattern used for the training, we want to take the average over the distribution of random binary patterns. While calculating $\langle Z \rangle_\xi$ would be relatively simple, calculating $\langle \log Z \rangle_\xi$ is much harder. Luckily we can use a technique called the \textit{replica trick}:
\begin{equation}\label{loglim}
\langle \log Z \rangle_\xi = \lim_{n\to 0} \frac{\langle Z^n \rangle_\xi -1}{n}.
\end{equation}
We have $n\in \mathbb N$, but treating it as a continuous variable works for the limit \cite{SantaFe}. In that case $Z^n$ is the partition function of $n$ replicas of the system:
\begin{equation}\label{zn}
\langle Z^n \rangle_\xi = e^{-\beta pn/2} \left \langle \Tr_S  \prod_{\rho=1}^n\int\prod_{\mu=1}^p dm^\mu_\rho \, \sqrt{\frac{\beta N}{2\pi}} \exp\left(-\frac{1}{2}\beta N (m_\rho^\mu)^2+ \beta m_\rho^\mu \sum_{i=1}^N \xi_i^\mu S_i^\rho \right)\right\rangle_\xi.
\end{equation}
Note that the average $\langle \hdots \rangle_\xi$ is over the same $Np$ variables $\xi_i^\mu$ as the previous case.
\subsection{Condensed pattern states} \label{h4b}
For $\alpha\neq 0$, in the limit $N\to \infty$ we will also have $p \to \infty$. We will focus on states that have a nonzero overlap with only a finite number $s$ of the $p$ stored patterns, called condensed patterns, that being equivalent to assuming $m_\rho^{\mu>s}\ll 1$ for the saddle point values. We now evaluate the last term in the integrand for a fixed $\mu > s$, using the approximation $\log \cosh x=x^2/2$ for small $x$:\\
\begin{equation}
\begin{aligned}
 \left \langle\prod_{\rho=1}^n \exp\left(\beta m_\rho^\mu \sum_{i=1}^N \xi_i^\mu S_i^\rho \right)\right\rangle_\xi &=  \prod_{i=1}^N\left \langle \exp\left(\beta  \xi_i^\mu\sum_{\rho=1}^nm_\rho^\mu S_i^\rho \right)\right\rangle_\xi =\\
 &=  \prod_{i=1}^N \cosh \left(\beta \sum_{\rho=1}^nm_\rho^\mu S_i^\rho \right )=\\
 &=\exp\left(\sum_{i=1}^N\log\cosh\left(\beta \sum_{\rho=1}^nm_\rho^\mu S_i^\rho\right)\right)\\
 &\simeq \exp\left(\frac{\beta^2}{2}\sum_{i=1}^N\left( \sum_{\rho=1}^nm_\rho^\mu S_i^\rho\right)^2\right)=\\
 &=\exp\left(\frac{\beta^2}{2}\sum_{i=1}^N\sum_{\rho=1}^n\sum_{\sigma=1}^nm_\rho^\mu m_\sigma^\mu S_i^\rho S_i^\sigma\right).
\end{aligned}
\end{equation}
We now introduce the matrix $\tilde \Lambda$:
\begin{equation}
\tilde \Lambda_{\rho\sigma}=\delta_{\rho\sigma}-\frac{\beta}{N} \sum_{i=1}^NS_i^\rho S_i^\sigma\,.
\end{equation}
So, for a $\mu >s$ we have that the integrand in equation \ref{zn} becomes:
\begin{equation}
E:=\left \langle \exp\left(-\frac{1}{2}\beta N (m_\rho^\mu)^2+ \beta m_\rho^\mu \sum_{i=1}^N \xi_i^\mu S_i^\rho \right)\right\rangle_\xi = \exp\left(\frac{\beta N}{2} \sum_{\rho, \sigma =1}^N\tilde \Lambda_{\rho\sigma} m^\mu_\rho m^\mu_{\sigma}\right).
\end{equation}
We can now evaluate the $n$-dimensional Gaussian integral:
\begin{equation}
\int \left(\prod_{\rho=1}^n dm^\mu_\rho \, \sqrt{\frac{\beta N}{2\pi}}\right) E = 
\left(\frac{\beta N}{2\pi}\right)^{n/2}\sqrt{\frac{\pi^n}{\det(\frac{1}{2}\beta N \tilde \Lambda)}}=(\det \tilde \Lambda)^{-1/2}.
\end{equation}
This is the contribution we get for every $\mu > s$. Since $s$ is kept finite, we have slightly less than $p$ contributions like this. Therefore the overall contribution is about:
\begin{equation}
(\det \tilde \Lambda)^{-p/2}=\exp \left(-\frac{p}{2}\log\det \tilde \Lambda\right)=\exp \left(-\frac{p}{2}\log\prod_{\rho=1}^n \tilde \lambda _\rho \right)=\exp \left(-\frac{p}{2}\sum_{\rho=1}^n\log \tilde \lambda _\rho \right),
\end{equation}
where $\tilde \lambda_\rho$, for $\rho=1, \hdots, n$ are the eigenvalues of $\tilde \Lambda$.\\
Our aim is to take the trace, which means that this term, togheter with the one for $\mu \leq s$ should be summed over the possible values of the $S^\rho_i$. A  very hard task, since the $S$-dependence is now buried in the eigenvalues $\tilde \lambda_\rho$. We therefore need more auxiliary variables. Let us start by defining $\Lambda$, a generalization of the matrix $\tilde \Lambda$ in which we introduce the parameters $q_{\rho\sigma}$:
\begin{equation}
\Lambda_{\rho\sigma}=(1-\beta)\delta_{\rho\sigma}-\beta q_{\rho\sigma}.
\end{equation}
We have $\Lambda = \tilde \Lambda$ for:
\begin{equation}
q_{\rho\sigma}=\left\{\!\! \begin{array}{ll} N^{-1}\sum_i S_i^\rho S_i^\sigma, & \rho\neq \sigma,\\ 0,& \text{otherwise.}
\end{array}\right.
\end{equation}
The value of any function $G(\tilde \lambda_1, \hdots, \lambda_n)$ of the eigenvalues of $\tilde \Lambda$ can be related to its value on the eigenvalues of $\Lambda$ using a Dirac delta:
\begin{equation}
G\{\tilde \lambda_{\rho}\}=\int \left(\prod_{(\rho, \sigma)}dq_{\rho\sigma}\delta\left(q_{\rho\sigma} - \frac{1}{N}\sum_{i=1}^N S_i^\rho S_i^\sigma\right)\right)G\{\lambda_\rho\},
\end{equation}
where the product extends over the distinct pairs $(\rho, \sigma)$. \\
In order to express the Dirac delta as an integral, 
\begin{equation}
\delta(x)=\frac{N\alpha \beta^2}{2\pi i}  \int_{-i\infty}^{+i\infty}e^{N\alpha\beta^2 rx}\, dr,
\end{equation}
we introduce another set of auxiliary variables $r_{\rho\sigma}$. We use this representation for every delta, that is $n(n-1)/2$ times:
\begin{equation}
G\{\tilde \lambda_{\rho}\}=\left(\frac{N\alpha \beta^2}{2\pi i}\right)^{\!n(n-1)/2}\int \left(\prod_{(\rho, \sigma)}dq_{\rho\sigma}dr_{\rho\sigma}\exp \left(N\alpha\beta^2 r_{\rho\sigma} q_{\rho\sigma} - \alpha\beta^2 r_{\rho\sigma}\sum_{i=1}^N S_i^\rho S_i^\sigma\right)\right)G\{\lambda_\rho\},
\end{equation}
We now apply this transformation to $\langle Z^n\rangle_\xi$, leaving out for semplicity all the uninportant prefactors:
\begin{equation}\label{zn2}\begin{aligned}
\langle Z^n \rangle_\xi &\propto \int \left(\prod_{\mu=1}^s\prod_{\rho=1}^n dm^\mu_\rho\right)\left(\prod_{(\rho,\sigma)}dq_{\rho\sigma}dr_{\rho\sigma}\right) \\
&\times \exp \left(-\frac{\beta N}{2} \sum_{\mu=1}^s\sum_{\rho=1}^n (m^\mu_\rho)^2-\frac{\alpha N}{2}\sum_{\rho=1}^n \log \lambda_\rho - \frac{N\alpha\beta^2}{2}\sum_{\rho=1}^n \sum_{\sigma \neq \rho} r_{\rho\sigma} q_{\rho\sigma}\right) \\
&\times \left \langle \Tr_S  \exp\left(\beta \sum_{\mu=1}^sm_{\rho}^\mu\sum_{i=1}^N \xi_i^\mu S_i^\rho + \frac{\alpha \beta^2}{2}\sum_{i=1}^N\sum_{\rho=1}^n \sum_{\sigma \neq \rho} r_{\rho\sigma}S_i^\rho S_i^\sigma \right)\right\rangle_\xi.
\end{aligned}
\end{equation}
Note that the sum over distinct pairs $(\sigma, \rho)$ has become half a sum over $\sigma$ and $\rho$ with $\rho \neq \sigma$.\\
The last line of equation \eqref{zn2} is the $\xi$ average of an expression like:
\begin{equation}
X=\Tr_S \exp\left(\sum_{i=1}^N F\{S_i,\xi_i\}\right)=\prod_{i=1}^N \Tr_{S_i}\exp F\{S_i,\xi_i\} =
\exp\left(\sum_{i=1}^N \log \Tr_{S_i} \exp F\{S_i,\xi_i\} \right),
\end{equation}
where $F$ is calculated on one site at a time, and therefore depends on the condensed patterns $\xi_i^1, \hdots, \xi_i^s$ and on the replica values of the site $S_i^1, \hdots S_i^n$. The trace $\Tr_S$ is on all $i$'s and $\rho$'s while $\Tr_{S_i}$ is just on the replicas index $\rho$. \\
Since every site can assume the same values $\pm 1$, the values of the trace $\Tr_{S_i} \exp F\{S_i,\xi_i\}$ would be the same for all $i$'s, except for the dependence on $\xi_i^\mu$. Nevertheless, in the limit $N\to \infty$, the number of sites is much larger than the number of possible sets $\{\xi_i^\mu\}_{\mu=1}^s$ at fixed $i$, that is $N\gg 2^s$, and therefore the sum over $i$ is equivalent to an average over patterns: $\langle \hdots \rangle_\xi = \frac{1}{N}\sum_{i=1}^N$ , the same self averaging introduced in \eqref{xiavg}.
\begin{equation}
X= \exp\left(N \left \langle \log \Tr_{S} \exp F\{S,\xi\}\right\rangle_\xi \right),
\end{equation}
the $i$-indices have disappeared and we have in effect a single unit with $n$ different replicas of $S^\rho$ and $p$ different $\xi^\mu$, as we can expect from a mean field method. We can also get rid of the outer average $\langle \hdots \rangle_\xi$ in the last line of \eqref{zn2} because the inner one already does the job.
We can thus write:
\begin{equation}\label{zn3}
\langle Z^n \rangle_\xi \propto \int \left(\prod_{\mu=1}^s\prod_{\rho=1}^n dm^\mu_\rho\right)\left(\prod_{(\rho,\sigma)}dq_{\rho\sigma}dr_{\rho\sigma}\right) e^{-N\beta f\{m,q,r\}}, 
\end{equation}
where:
\begin{equation}\label{f}
\begin{aligned}
f\{m,q,r\}&=\frac{1}{2} \sum_{\mu=1}^s\sum_{\rho=1}^n (m^\mu_\rho)^2+\frac{\alpha}{2\beta}\sum_{\rho=1}^n \log \lambda_\rho + \frac{\alpha\beta}{2}\sum_{\rho=1}^n \sum_{\sigma \neq \rho} r_{\rho\sigma} q_{\rho\sigma}+\\
&-\frac{1}{\beta}\left\langle \log \Tr_S\exp \left(\beta\sum_{\mu=1}^s\sum_{\rho=1}^nm^\mu_\rho\xi^\mu S^\rho+\frac{\alpha\beta^2}{2}\sum_{\rho=1}^n \sum_{\sigma \neq \rho} r_{\rho\sigma} S^{\rho} S^\sigma \right)\right\rangle_\xi.
\end{aligned}
\end{equation}
For large $N$ we can obtain the free energy per unit using the $\log$ limit \eqref{loglim} and the saddle point method on $\int e^{-\beta N f}$:
\begin{equation}\label{FN}
\begin{aligned}
F/N=-\frac{1}{\beta N}\langle \log Z \rangle_\xi&=-\frac{1}{\beta N}\lim_{n\to 0}\frac{1}{n}\left(\langle Z^n\rangle_\xi -1\right)=\\
&=-\frac{1}{\beta N}\lim_{n\to 0}\frac{1}{n} \log \langle Z^n \rangle_\xi=\\
&=\frac{\alpha}{2}+\lim_{n\to 0}\frac{1}{n} \min f\{m,q,r\}.
\end{aligned}
\end{equation}
The saddle point can be found extremizing $f$:
\begin{equation}\label{saddlepoint2}
\frac{\partial f}{\partial m_\rho^\mu}=0, \qquad\frac{\partial f}{\partial q_{\rho\sigma}}=0, \qquad\frac{\partial f}{\partial r_{\rho\sigma}}=0.
\end{equation}
In order to proceed we need to make an \textit{ansatz}, by assuming the \textbf{replica symmetry}. With this assumption, the meaning of the mean field parameters is now clear:
\begin{itemize}
\item $m^\mu=\frac{1}{N}\sum_i \xi_i^\mu\langle S_i\rangle_\beta$ is, as in the case $\alpha=0$, the overlap between the $\mu$-th pattern, $\bm \xi ^\mu$, and the network configuration $\bm S$.
\item $q=\langle \frac{1}{N}\sum_i \langle S_i\rangle^2_\beta\rangle_\xi$ is the mean squared magnetization.
\item $\alpha r =\sum_{\mu>s}\langle (m^\mu)^2 \rangle_\xi$ is the mean squared overlap with the \textit{uncondensed} patterns ($\mu>s$).
\end{itemize}
The function $f$ greatly simplifies to:
\begin{equation}\label{f5}
\begin{aligned}
f(\bm m, q, r) &=\frac{1}{2}n\bm m^2+\frac{\alpha}{2\beta}\sum_{\rho=1}^n \log \lambda_\rho + \frac{1}{2}\alpha\beta n(n-1) rq+\frac{1}{2}\alpha \beta n r \, +\\
&-\frac{1}{\beta}\left\langle \log \Tr_S\exp \left(\beta \bm m \cdot \bm \xi \sum_{\rho=1}^nS^\rho+ \frac{1}{2}\alpha \beta^2 r \left( \sum_{\rho=1}^n S^{\rho} \right)^{\!\!2}\right)\right\rangle_\xi.
\end{aligned}
\end{equation}
The matrix $\Lambda$ simplifies to:
\begin{equation}
\Lambda_{\sigma \rho} =\left\{\!\! \begin{array}{ll} 1-\beta , & \rho= \sigma,\\ -\beta q,& \rho \neq \sigma.
\end{array}\right.
\end{equation}
thus calculating the eigenvalues $\lambda_\rho$ is now much easier. They are found to be:
\begin{equation}
\lambda_1=1-\beta-(n-1)\beta q, \quad \lambda_2=\hdots=\lambda_n = 1-\beta(1-q), \end{equation}
and the sum of the $\log$'s:
\begin{equation}\begin{aligned}
\frac{1}{n} \sum_{\rho=1}^n \log\lambda_\rho &= \frac{1}{n} \left(\log(1-\beta-(n-1)\beta q)+(n-1)\log(1-\beta(1-q)) \right)\\
& \xrightarrow{n\to 0} \log(1-\beta(1-q))-\frac{\beta q}{1-\beta(1-q)}
\end{aligned}\end{equation}
We now need to evaluate the last term in \eqref{f5}, in particular, the trace over the $S$'s. To linearize the term in $S^2$ we use once again the Gaussian integral trick, introducing a new auxiliary variable $z$: 
\begin{equation}
\exp\left(\frac{1}{2}\alpha \beta^2r\left(\sum_{\rho=1}^nS^\rho\right)^2\right) = \int \frac{dz}{\sqrt{2\pi}} \exp \left(-\frac{1}{2}z^2+\beta\sqrt{\alpha r} z \sum_{\rho=1}^nS^\rho\right). \end{equation}
Giving:
\begin{equation}\begin{aligned}
Y&= \Tr_S\exp \left(\beta \bm m \cdot \bm \xi \sum_{\rho=1}^nS^\rho+ \frac{1}{2}\alpha \beta^2 r \left( \sum_{\rho=1}^n S^{\rho} \right)^{\!\!2}\right) =\\
&= \Tr_S  \int \frac{dz}{\sqrt{2\pi}} \exp \left(-\frac{1}{2}z^2+\beta(\sqrt{\alpha r} z + \bm m\cdot \bm \xi)\sum_{\rho=1}^nS^\rho\right) =\\
&= \int \frac{dz}{\sqrt{2\pi}} e^{-z^2/2} \left(2 \cosh\beta(\sqrt{\alpha r} z + \bm m\cdot \bm \xi)\right)^n=\\
&= \int \frac{dz}{\sqrt{2\pi}} e^{-z^2/2} \exp\left(n \log 2\cosh\beta(\sqrt{\alpha r} z + \bm m\cdot \bm \xi)\right)
\end{aligned}\end{equation}
We now take the $\log$, average over patterns, multiply for $1/n$ and take the limit $n\to 0$. We can therefore expand for small $n$'s:
\begin{equation}\begin{aligned}
\frac{1}{n}\langle \log Y \rangle_\xi &\simeq \frac{1}{n}\left\langle \log  \int \frac{dz}{\sqrt{2\pi}} e^{-z^2/2} \left(1 + n \log 2\cosh\beta(\sqrt{\alpha r} z + \bm m\cdot \bm \xi)\right)\right\rangle_\xi \simeq \\
&\simeq  \frac{1}{n}\left\langle n  \int \frac{dz}{\sqrt{2\pi}} e^{-z^2/2}  \log \left(2\cosh\beta(\sqrt{\alpha r} z + \bm m\cdot \bm \xi)\right)\right\rangle_\xi\\
&\xrightarrow{n\to 0}  \left\langle \log \left(2\cosh\beta(\sqrt{\alpha r} z + \bm m\cdot \bm \xi)\right)\right\rangle_{\xi,z},
\end{aligned}\end{equation}
Where the average $\langle \hdots \rangle_z$ means an average over the Gaussian random field given by the effects of the \textit{uncondensed patterns} $\mu >s$.
Putting this results in equation \eqref{FN} yields the expression for the free energy per site presented in \eqref{fepu}, which gives the saddle point equations \eqref{sys}.
\newpage
 \subsection{Solution for pure states}
 \label{h5}
We now solve the self consistency equations \eqref{sys} for pure states, $\bm m = (m, 0, \hdots, 0)$.
In this case, the system simplifies to:
\begin{equation}\begin{cases}
m=\langle \tanh\beta(z\sqrt{\alpha r}+m)  \rangle_{z}\, ,\\
q=\langle \tanh^2\beta(z\sqrt{\alpha r}+m)  \rangle_{z}\,,\\
r=\frac{q}{(1-\beta(1-q))^2}.
\end{cases}\end{equation}
We solved it numerically using a python code for some values of $(\alpha, T) \in [0,\,0.15] \times[0,\,1]$.We found that for $\alpha \lesssim 0.10$ the $m$ parameter undergoes a double \textbf{phase transition} (I and II order), while for bigger values just a single one (I order).
We can therefore identify 3 different regions in the plane, the bottom 2 correspond to stable memory state, while the upper one to unstable states.
\begin{figure}[h]
\vspace{0.5cm}
\caption{Pure state solutions to the self consistent equations \eqref{sys}}
     \centering
     \begin{subfigure}[l]{0.6\textwidth}
       \centering
	\includegraphics[width=\textwidth]{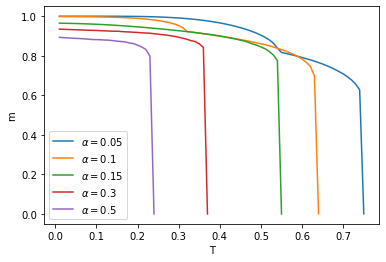}
		\captionsetup{width=\textwidth}	
	\caption{\textit{Magnetization} mean field parameter $m$ values for $T \in [0,\,1]$ as given 	by the system \eqref{sys} for fixed $\alpha$'s.}
	\label{mfalfa}
     \end{subfigure}
     \quad
     \begin{subfigure}[r]{0.3\textwidth}
	\centering
	\captionsetup{width=\textwidth}	
	\includegraphics[width=\textwidth]{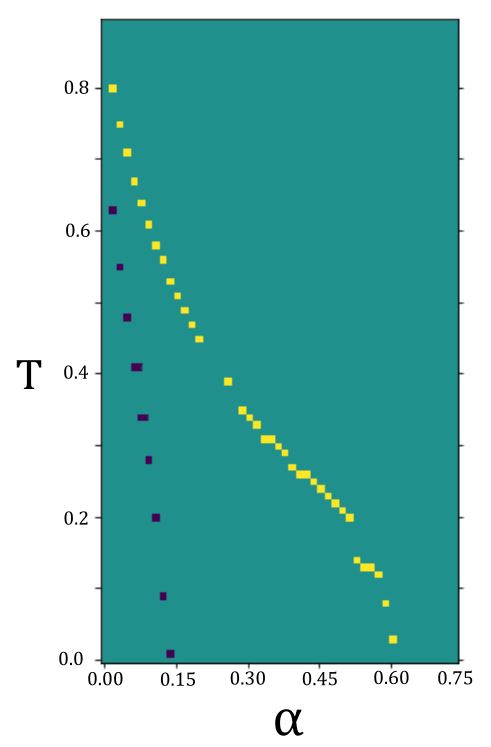}
	\caption{Phase diagram of the system.}
	\label{phasediag}
     \end{subfigure}
\end{figure}
\\
The equations have been numerically solved using a python code.
We first defined the integral functions using the \textit{quad} function from \textit{scipy} module.
\vspace{1cm}
\begin{lstlisting}[frame=single, language=Python] 
from scipy.integrate import quad
import numpy as np

def g1(x, r, m, a, t):
    return np.tanh((np.sqrt(a*r)*x + m)/t)*
    			np.exp(-x**2/2)/np.sqrt(2*np.pi)

def g2(x, r, m, a, t):
    return np.tanh((np.sqrt(a*r)*x + m)/t)**2*
    			np.exp(-x**2/2)/np.sqrt(2*np.pi)

def f1(r,m,a,t):
    return quad(g1, -np.inf, np.inf, args=(r,m,a,t))[0] - m

def f2(r,m,q,a,t):
    return quad(g2, -np.inf, np.inf, args=(r,m,a,t))[0] - q

def f3(r,q,t):
    return q/(1 - (1 - q)/t)**2 - r
\end{lstlisting} 
\vspace{1cm}

And solved them, for the values of $\alpha = 0.015 j$, with $j=1, \hdots, 50$ and $T=0.01 k$ 
with $k=1, \hdots, 90$, using the \textit{fsolve} function:
\newpage
\begin{lstlisting}[frame=single, language=Python] 
from scipy.optimize import fsolve

def func(x,a,t):
    return [f1(x[0],x[1],a,t), f2(x[0],x[1], x[2],a,t), 
    			f3(x[0],x[2],t) ]
    
## Phase Diagram
imax=50
jmax=90
aT= np.zeros((imax,jmax))

for i in range(imax):
    a=(i+1)*0.015
    for j in range(jmax):
        T=(j+1)*0.01
        m=fsolve(func, [1,1,1], args=(a,T))[1]
        if(m<0.01):
            break
        else:
            aT[i,j]=m
\end{lstlisting} 
\vspace{1cm}

In order to identify the phase transitions, we evaluated the numerical derivatives with the respect to $T$ of the solutions $m(T)$, applied an high-pass filter and plotted the results as the pixel plot \ref{phasediag}.

 \newpage
  \section{Gardner's theory}
  \subsection{Simple perceptron}
 \label{h6}
 The fraction of volume in weight space obtained in \eqref{volume} looks like a partition function implementing an all-or-nothing $\Theta$ distribution \cite{SantaFe} whose statistical properties are embodied by the the \textit{free energy} $G=\frac{1}{N}\langle \log V\rangle_{\xi, \zeta}$. While averaging $V$ over the $\xi$'s and $\zeta$'s would be relatively simple, doing so with $\log V$ is much more complicated. Therefore, we will need to apply the replica trick introduced in Section \ref{h4b}:
 \begin{equation} \label{rep2}
NG=\langle \log V \rangle_{\zeta,\xi} = \lim_{n\to 0} \frac{\langle V^n\rangle_{\zeta,\xi} -1}{n}.
\end{equation}
First, we analyze the expression of $V$, the volume of a single replica obtained in \eqref{volume}, noticing that the index $i$ can be dropped: in fact expression \eqref{volume} factorizes into the product of $N$ identical terms, we can therefore reduce our calculations to a single output unit without loss of generality.
\begin{equation}\label{solid2} V=\frac{\int\prod_{j=1}^Ndw_{j} \, \prod_{\mu=1}^p\Theta(\zeta_i^\mu N^{-1/2}\sum_{j=1}^N w_{j} \xi_j-\kappa)\, \delta(\sum_{j=1}^Nw_{j}^2-N)}{\int\prod_{j=1}^Ndw_{j} \, \delta(\sum_{j=1}^Nw_{j}^2-N)}\,. \end{equation}
From this expression, we can write the one for the average of $n$ replicas, $\langle V^n\rangle_{\zeta,\xi}$: 
 \begin{equation} \label{rep21}
\langle V^n\rangle_{\zeta,\xi} = \frac{\left\langle \prod_{\rho=1}^n \int d\bm w^\rho\left(\prod_{\mu=1}^p\,\Theta\left(\zeta^\mu N^{-1/2}\sum_{j=1}^Nw_j^\rho\xi_j^\mu - \kappa\right)\right)\,\delta\left(\sum_{j=1}^N(w_j^\rho)^2-N\right)\right\rangle_{\xi,\zeta}}{\prod_{\rho=1}^n\int d\bm w^\rho \, \delta\left(\sum_{j=1}^N(w_j^\rho)^2-N\right)} \,.
\end{equation}
To proceed, we use the integral representation of the $\Theta$ function:
 \begin{equation} \label{Theta}
\Theta(z-\kappa)=\int_\kappa^\infty \delta(\lambda-z)=\int_\kappa^\infty d\lambda \int \frac{dx}{2\pi} e^{ix(\lambda-z)}.
\end{equation}
Given $z^\mu_\rho:=\zeta^\mu N^{-1/2}\sum_{j=1}^Nw_j^\rho \xi^\mu_j$, we have step functions for each $\mu$ and $\rho$, so we need auxiliary variables $x^\mu_\rho$ and $\lambda^\mu_\rho$:
 \begin{equation} \label{Theta2}
\Theta\left(\zeta^\mu N^{-1/2}\sum_{j=1}^Nw_j^\rho \xi^\mu_j-\kappa\right)=\int_\kappa^\infty d\lambda^\mu_\rho \int \frac{dx^\mu_\rho}{2\pi}\, e^{ix^\mu_\rho\lambda^\mu_\rho}e^{-ix^\mu_\rho z^\mu_\rho}.
\end{equation}
The patterns are now factorized apart (they occur only in the last factor) so it is easy to evaluate the average $\langle \hdots \rangle_{\xi,\zeta}$. Moreover, we are considering independent binary patterns, so the average is essentialy equivalent to taking half $\zeta^\mu \xi_j^\mu=+1$ and half $\zeta^\mu \xi_j^\mu=-1$ :
\begin{equation}
\begin{aligned}
\left \langle \prod_{\mu=1}^p\prod_{\rho=1}^n e^{-ix^\mu_\rho z^\mu_\rho}\right\rangle &=
 \prod_{\mu=1}^p\prod_{j=1}^N  \left \langle \exp \left(-i\zeta^\mu \xi^\mu_j N^{-1/2}\sum_{\rho=1}^nw_j^\rho x^\mu_\rho \right)\right\rangle=\\
 &=\exp\left(\sum_{\mu=1}^p\sum_{j=1}^N \log \cos \left(N^{-1/2}w_j^\rho x^\mu_\rho \right)\right)\\
 &\xrightarrow{N\to \infty} \exp\left(-\frac{1}{2N}\sum_{\mu=1}^p\sum_{\rho,\sigma=1}^n x^\mu_\rho x^\mu_\sigma\sum_{j=1}^Nw_j^\rho w_j^\sigma\right).
\end{aligned}
\end{equation}
In the last line we used $\log \cos x \simeq \frac{x^2}{2}$. We introduce a \textit{correlation} variable to represent the last term:
\begin{equation}\label{qrs} q_{\rho\sigma}=\frac{1}{N}\sum_{j=1}^Nw_j^\rho w_j^\sigma.\end{equation}
To enforce this condition, we introduce in \eqref{solid2} a $\delta\big(q_{\rho\sigma}-\frac{1}{N}\sum_{j=1}^Nw_j^\rho w_j^\sigma\big)$ and an integral over the $q_{\rho\sigma}$'s.
From the normalization of the weights \eqref{norm}, it is clear that $q_{\rho\rho}=1$, $\forall \rho=1,\hdots,n$. Moreover, we have $q_{\rho\sigma}=q_{\sigma\rho}$. We can thus split the sum in two:
\begin{equation}
\begin{aligned}
\left \langle \prod_{\mu=1}^p\prod_{\rho=1}^n e^{-ix^\mu_\rho z^\mu_\rho}\right\rangle &=
\prod_{\mu=1}^p \exp\left(-\frac{1}{2}\sum_{\rho=1}^n (x^\mu_\rho)^2-\sum_{\rho=1}^n \sum_{\sigma<\rho}q_{\rho\sigma}x^\mu_\rho x^\mu_\sigma\right).
\end{aligned}
\end{equation}
Since we get this same result for each of the $p$ $\Theta$ function, we can drop the $\mu$ index:
\begin{equation}\label{1111}
\begin{aligned}
\left \langle \prod_{\mu=1}^p\prod_{\rho=1}^n \Theta(z^\mu_\rho-\kappa)\right\rangle &=
\left (\int_\kappa^\infty \left(\prod_{\rho=1}^n\frac{d\lambda_\rho}{2\pi}\right)\int_{-\infty}^{+\infty}\left(\prod_{\rho=1}^n dx_\rho\right)e^{K\{\lambda,x,q\}}\right)^p.
\end{aligned}
\end{equation}
We also defined:
\begin{equation}
K\{\lambda,x,q\}=i \sum_{\rho=1}^nx_\rho\lambda_\rho-\frac{1}{2}\sum_{\rho=1}^nx_\rho^2- \sum_{\rho=1}^n\sum_{\sigma < \rho}q_{\rho\sigma}x_\rho x_\sigma.
\end{equation}
It is now time to deal with the $\delta$'s, using the integral representation:
\begin{equation}
\delta(z)=\int_{-i\infty}^{+i\infty} \frac{dr}{2\pi i} e^{-rz}.
\end{equation}
To treat the $\delta\left(\sum_{j=1}^N(w_j^\rho)^2-N\right)$, which comes from the normalization \eqref{norm}, we choose $r=E_\rho/2$, obtaining:
\begin{equation}
\delta\!\left(\sum_{j=1}^N(w_j^\rho)^2-N\right)=\int_{-i\infty}^{+i\infty} \frac{dE_\rho}{4\pi i} e^{\frac{N}{2}E_\rho-\frac{1}{2}E_\rho\sum_{j=1}^N(w_j^\rho)^2}.
\end{equation}
For the $\delta\big(q_{\rho\sigma}-\frac{1}{N}\sum_{j=1}^Nw_j^\rho w_j^\sigma\big)$ coming from \eqref{qrs} we instead use $r=NF_{\rho\sigma}$:
\begin{equation}
\delta\!\left(q_{\rho\sigma}-\frac{1}{N}\sum_{j=1}^Nw_j^\rho w_j^\sigma\right)=N\int_{-i\infty}^{+i\infty} \frac{dF_{\rho\sigma}}{2\pi i} e^{-NF_{\rho\sigma}q_{\rho\sigma}+ F_{\rho\sigma}\sum_{j=1}^N w_j^\rho w_j^\sigma}.
\end{equation}
We can now factorize the integral over $w_j^\rho$ in equation \eqref{rep21}. Moreover, the index $j$ is now a dummy index, and can be dropped:
 \begin{equation}\label{fact}
 \begin{aligned}
&\int \left(\prod_{j=1}^N dw^\rho_j\right) \delta\!\left(q_{\rho\sigma}-\frac{1}{N}\sum_{j=1}^Nw_j^\rho w_j^\sigma\right)\delta\!\left(\sum_{j=1}^N(w_j^\rho)^2-N\right)=\\
=&N\int_{-i\infty}^{+i\infty} \frac{dE_\rho}{4\pi i} \int_{-i\infty}^{+i\infty} \frac{dF_{\rho\sigma}}{2\pi i}\,e^{\frac{N}{2}E_\rho -NF_{\rho\sigma}q_{\rho\sigma}}
\int\left(\prod_{j=1}^N dw^\rho_j\right)\,e^{-\frac{1}{2}E_\rho\sum_{j=1}^N(w_j^\rho)^2 
  +F_{\rho\sigma}\sum_{j=1}^N w_j^\rho w_j^\sigma}=\\
=&N\int_{-i\infty}^{+i\infty} \frac{dE_\rho}{4\pi i} \int_{-i\infty}^{+i\infty} \frac{dF_{\rho\sigma}}{2\pi i}\,e^{\frac{N}{2}E_\rho-NF_{\rho\sigma}q_{\rho\sigma}}
 \left(\int dw^\rho\, e^{-\frac{1}{2}E_\rho (w^\rho)^2 
  +F_{\rho\sigma} w^\rho w^\sigma}\right)^N=\\
  =&N\int_{-i\infty}^{+i\infty} \frac{dE_\rho}{4\pi i} \int_{-i\infty}^{+i\infty} \frac{dF_{\rho\sigma}}{2\pi i}\,\exp\left(\frac{N}{2}E_\rho-NF_{\rho\sigma}q_{\rho\sigma}+\right.+\\
  +&\left.N\log \int \!dw^\rho \exp\left(-\frac{1}{2}E_\rho (w^\rho)^2 
  +F_{\rho\sigma} w^\rho w^\sigma\right)\right).
\end{aligned}
\end{equation}
Collecting all the factors together we get:
 \begin{equation} \label{rep3}
\langle V^n\rangle_{\zeta,\xi} = \frac{
\int \left(\prod_{\rho=1}^n dE_\rho\right)\left(\prod_{\rho=1}^n \prod_{\sigma<\rho} dF_{\rho\sigma}dq_{\rho\sigma}\right) e^{NG\{q,F,E\}}
}{
\int \left(\prod_{\rho=1}^n dE_\rho\right) e^{NH\{E\}}
},
\end{equation}
where (first line comes from the $\delta$'s, second one from the $\langle \Theta(\hdots)\rangle_\xi$):
 \begin{equation}\label{G}
 \begin{aligned}
G\{q,F,E\} =&\frac{1}{2}\sum_{\rho=1}^n E_\rho-\sum_{\rho=1}^n \sum_{\sigma<\rho} F_{\rho\sigma}q_{\rho\sigma}+\\
&+\log \int \!dw^\rho \exp\left(-\frac{1}{2}\sum_{\rho=1}^n E_\rho (w^\rho)^2 
  +\sum_{\rho=1}^n \sum_{\sigma<\rho}F_{\rho\sigma} w^\rho w^\sigma\right) +\\
  &+\frac{p}{N}\log \left (\int_\kappa^\infty \left(\prod_{\rho=1}^n\frac{d\lambda_\rho}{2\pi}\right)\int_{-\infty}^{+\infty}\left(\prod_{\rho=1}^n dx_\rho\right)e^{K\{\lambda,x,q\}}\right),
\end{aligned}
\end{equation}
and:
 \begin{equation}\label{hfunc}
H\{E\} =\frac{1}{2}\sum_{\rho=1}^n E_\rho+\log \int \!dw^\rho \exp\left(-\frac{1}{2}\sum_{\rho=1}^n E_\rho (w^\rho)^2 \right)
\end{equation}
Once again, the exponents inside the integrals are proportional to $N$, so we will use the saddle-point method in the limit of large $N$, and once again, we will assume replica symmetry:
 \begin{equation}\label{replica}
q_{\rho\sigma}=q, \qquad F_{\rho\sigma}=F, \qquad E_\rho=E.
\end{equation}
We can now study $G$. For the last term, we can write $K$ as:
\begin{equation}
K\{\lambda,x,q\}=i \sum_{\rho=1}^nx_\rho\lambda_\rho-\frac{1-q}{2}\sum_{\rho=1}^nx_\rho^2-\frac{q}{2}\left( \sum_{\rho=1}^n x_\rho\right)^2.
\end{equation}
The last term can be linearized with the usual Gaussian integral trick:
\begin{equation}
e^{-\frac{q}{2}\left(\sum_\rho x_\rho\right)^2}=\int \frac{dt}{\sqrt{2\pi}}e^{-t^2+it\sqrt{q}\sum_\rho x_\rho}
\end{equation}
The integrals over the $x_\rho$'s and $\lambda_\rho$'s now factorize, and we can also evaluate the one over $x$:
\begin{equation}\label{eq119}\begin{aligned}
\int \left(\prod_{\rho=1}^n  \frac{dx_\rho\,d\lambda_\rho}{2\pi}\right) \,e^{K\{\lambda,x,q\}}&=
\int \frac{dt}{\sqrt{2\pi}} e^{-t^2}\prod_{\rho=1}^n\int  \frac{dx_\rho\,d\lambda_\rho}{2\pi} \, e^{i(\lambda_\rho+\sqrt{q}t)x_\rho-\frac{1-q}{2}x_\rho^2}=\\
&=\int \frac{dt}{\sqrt{2\pi}} e^{-t^2}\left(\int  \frac{dx\,d\lambda}{2\pi} \, e^{i(\lambda+\sqrt{q}t)x-\frac{1-q}{2}x^2}\right)^n=\\
&=\int \frac{dt}{\sqrt{2\pi}} e^{-t^2} \left(\int_\kappa^\infty\frac{d\lambda}{\sqrt{2\pi(1-q)}}\exp\left(-\frac{(\lambda+t\sqrt{q})^2}{2(1-q)}\right)\right)^n.
\end{aligned}
\end{equation}
The whole last line of \eqref{G} is, for $\alpha:=p/N$, as usual:
\begin{equation}\label{eq120}\begin{aligned}
&\alpha \log\int \frac{dt}{\sqrt{2\pi}} e^{-t^2} \left(\int_\kappa^\infty\frac{d\lambda}{\sqrt{2\pi(1-q)}}\exp\left(-\frac{(\lambda+t\sqrt{q})^2}{2(1-q)}\right)\right)^n \\
&\xrightarrow{n\to 0} n\alpha
\int \frac{dt}{\sqrt{2\pi}} e^{-t^2} \log\int_\kappa^\infty\frac{d\lambda}{\sqrt{2\pi(1-q)}}\exp\left(-\frac{(\lambda+t\sqrt{q})^2}{2(1-q)}\right).
\end{aligned}
\end{equation}
The first line of \eqref{G} can be evaluated in a similar fashion, linearizing the term $(\sum_\rho w_\rho)^2$ with a Gaussian integral trick and then integrating over the $w_\rho$. The final result is:
\begin{equation}\label{eq121}\frac{1}{2}n\left(E+qF+\log(2\pi)-\log(E+F)+\frac{F}{E+F}\right).
\end{equation}
We can now minimize $G(q,E,F)$ to find the saddle point. From $\frac{\partial G}{\partial E}=0$ and $\frac{\partial G}{\partial F}=0$, we find, respectively:
\begin{equation}\label{eq122}F=\frac{q}{(1-q)^2}, \qquad E=\frac{1-2q}{(1-q)^2}.
\end{equation}
With these substitutions and a change of variable in the integral $\lambda \to z$, we can also evaluate $\frac{\partial G}{\partial q}=0$, obtaining:
\begin{equation}
\alpha \int \frac{dt}{\sqrt{2\pi}} e^{-t^2/2} \left(\int_u^\infty dz\, e^{-z^2/2}\right)^{-1}\!e^{-u^2/2}\frac{u}{2\sqrt{q}(1-q)}=\frac{q}{2(1-q)^2}.
\end{equation}
where $u=\frac{\kappa +t\sqrt{q}}{\sqrt{1-q}}$.\\ We now focus on the meaning of the parameter $q$: its saddle-point value is the most probable overlap value between pairs of solutions. When $\alpha$ is small, a large portion of the $w$-space solves the problem, therefore different solutions can be uncorrelated and thus $q \sim 0$; whereas when $\alpha$ reaches its maximum, all the solutions are \textit{packed together} in a small portion of $w$-space, and therefore highly correlated $q\to 1$. This is the condition for the \textit{optimal} perceptron, so taking $q\to 1$ yields eq.\! \eqref{alpha}, the expression of $\alpha(\kappa)$. As can be seen in fig. \!\!\!\! \ref{alfa}, for $\kappa=0$, we retrive the expected $\alpha=2$.

   \subsection{\textit{Quadratic} perceptron}
 \label{h7}
 In order to obtain the $G$ function that, through partial derivatives will give us the parameter values, we need to evaluate $\langle \log V\rangle_{\xi}$. To do so, we once again apply the replica trick. $\langle V^n\rangle_{\xi}$ can be calculated from \eqref{qsolid} :
 \begin{equation}\label{qsolid2} \langle V^n\rangle_{\xi}=\frac{\left \langle\prod_{\rho=1}^n\int\prod_{j=1}^Ndw^\rho_{j} \, \prod_{\mu=1}^p\Theta\left(N^{-1}\left\vert \sum_{j=1}^N w^\rho_{j}\, \xi_j^\mu\right\vert^2 -\kappa\right)\,\delta\left(\sum_{j=1}^N|w^\rho_{j}|^2-N\right)\right\rangle_{\xi}}{\prod_{\rho=1}^n \int\prod_{j=1}^Ndw^\rho_{j} \,\delta\left(\sum_{j=1}^N|w^\rho_{j}|^2-N\right)}\,, \end{equation}
where we have dropped the site index $i$ because, as we observed for the simple perceptron in \eqref{solid2}, $\prod_{i=1}^N$ is just a product of $N$ identical terms.
We now represent the $\Theta$ function using \eqref{Theta}. In particular, given $z^\mu_\rho = N^{-1}\left\vert \sum_{j=1}^N w^\rho_{j}\, \xi_j^\mu\right\vert^2$, we get:
 \begin{equation} \label{qTheta2}
\Theta\left(N^{-1}\left\vert \sum_{j=1}^N w^\rho_{j}\, \xi_j^\mu\right\vert^2-\kappa\right)=\int_\kappa^\infty d\lambda^\mu_\rho \int \frac{dx^\mu_\rho}{2\pi}\, e^{ix^\mu_\rho\lambda^\mu_\rho}e^{-ix^\mu_\rho z^\mu_\rho}.
\end{equation}
 It is now time to . In order to evaluate the average over the patterns $\langle \hdots\rangle_\xi$, we shall assume a an Ising-like input distribution. Observing that the $\xi$ arguments of the $\Theta$ now factorize, we find:
 \begin{equation}
\begin{aligned}
\left \langle \prod_{\mu=1}^p\prod_{\rho=1}^n e^{-ix^\mu_\rho z^\mu_\rho}\right\rangle &=
\prod_{\mu=1}^p\prod_{\rho=1}^n  \left \langle \exp \left(-\frac{i}{N}x_{\rho}^\mu\left\vert \sum_{j=1}^N w^\rho_{j}\, \xi_j^\mu\right\vert^2 \right)\right\rangle_{\xi}=\\
&=\prod_{\mu=1}^p\prod_{j=1}^N  \prod_{k=1}^N \left \langle \exp \left(-\frac{i}{N}\sum_{\rho=1}^n  x_{\rho}^\mu \, w^\rho_{j}w^\rho_{k}\, \xi_j^\mu\xi_k^\mu \right)\right\rangle_{\xi}=\\
&=\exp\left(\sum_{\mu=1}^p\sum_{j=1}^N\sum_{k=1}^N \log \cos \left(N^{-1} \sum_{\rho=1}^n w_j^\rho w_k^\rho x^\mu_\rho \right)\right)\\
&\xrightarrow{N\to \infty} \exp\left(-\frac{1}{2N^2}\sum_{\mu=1}^p\sum_{\rho,\sigma=1}^n x^\mu_\rho x^\mu_\sigma \sum_{j=1}^N \sum_{k=1}^Nw_j^\rho w_j^\sigma w_k^\rho w_k^\sigma\right).
\end{aligned}
\end{equation}
We also assumed that weights and patterns are real (second line) and that the weights correlate weakly (last line), so that we can Taylor-expand $\log\cos x\simeq \frac{x^2}{2}$. As in the previous case, eq. \eqref{qrs}, we introduce the $q_{\rho\sigma}$ variable, so:
\begin{equation}
\left \langle \prod_{\mu=1}^p\prod_{\rho=1}^n e^{-ix^\mu_\rho z^\mu_\rho}\right\rangle \simeq \exp\left(-\frac{1}{2N^2}\sum_{\mu=1}^p\sum_{\rho,\sigma=1}^n x^\mu_\rho x^\mu_\sigma q_{\rho\sigma}^2\right),
\end{equation}
As in \eqref{1111} we observe that we get this same result for each of the $p$, $\mu$-indexed, $\Theta(z^\mu_\rho-\kappa)$ functions, and we can therefore drop the $\mu$ index in favour of a $p$ exponent.
\begin{equation}
\begin{aligned}
\left \langle \prod_{\mu=1}^p\prod_{\rho=1}^n \Theta(z^\mu_\rho-\kappa)\right\rangle &=
\left (\int_\kappa^\infty \left(\prod_{\rho=1}^n\frac{d\lambda_\rho}{2\pi}\right)\int_{-\infty}^{+\infty}\left(\prod_{\rho=1}^n dx_\rho\right)e^{K\{\lambda,x,q\}}\right)^p,
\end{aligned}
\end{equation}
where:
 \begin{equation}\label{eqK}
K\{\lambda,x,q\}=i \sum_{\rho=1}^nx_\rho\lambda_\rho-\frac{1}{2}\sum_{\rho=1}^nx_\rho^2- \sum_{\rho=1}^n\sum_{\sigma < \rho}(q_{\rho\sigma})^2x_\rho x_\sigma.
\end{equation}
The $q^2$ in this function will constitute the only difference between this case and the standard simple perceptron.\\
Introducing the $q_{\rho\sigma}$ variable came with the cost of introducing also a $\delta\left(q_{\rho\sigma}- \frac{1}{N}\sum_{j=1}^N w_j^\rho w_j^\sigma \right)$ and an integral over these $q$'s.  These $\delta$'s will give the exact same contribution of the previous case. In fact, using the integral representation we have:
\begin{equation} \begin{aligned}
&\delta\left(\sum_{j=1}^N(w_j^\rho)^2-N\right) =\int_{-i\infty}^{+i\infty} \frac{dE_\rho}{2\pi i} \,\exp \left(E_\rho\left(\sum_{j=1}^N(w_j^\rho)^2-N\right)\right), \\
& \delta\left(q_{\rho\sigma}- \frac{1}{N}\sum_{j=1}^N w_j^\rho w_j^\sigma  \right)= N
\int_{-i\infty}^{+i\infty} \frac{dF_{\rho\sigma}}{2\pi i} \, \exp \left(NF_{\rho\sigma}\left(q_{\rho\sigma}- \frac{1}{N}\sum_{j=1}^N w_j^\rho w_j^\sigma  \right)\right)
\end{aligned}\end{equation}
And,as in \eqref{fact}, we can now factorize the integral over $w_j^\rho$ in equation \eqref{qsolid2}. Once again, the index $j$ becomes a dummy index that can be dropped:
 \begin{equation}
 \begin{aligned}
&\int \left(\prod_{j=1}^N dw^\rho_j\right) \delta\!\left(q_{\rho\sigma}-\frac{1}{N}\sum_{j=1}^Nw_j^\rho w_j^\sigma\right)\delta\!\left(\sum_{j=1}^N(w_j^\rho)^2-N\right)=\\
  =&N\int_{-i\infty}^{+i\infty} \frac{dE_\rho}{2\pi i} \int_{-i\infty}^{+i\infty} \frac{dF_{\rho\sigma}}{2\pi i}\,\exp\left(\frac{N}{2}E_\rho-NF_{\rho\sigma}q_{\rho\sigma}+\right.\\
  +&\left.N\log \int \!dw^\rho \exp\left(-\frac{1}{2}E_\rho (w^\rho)^2 
  +F_{\rho\sigma} w^\rho w^\sigma\right)\right),
\end{aligned}
\end{equation}
And we are left with the same contribution as  \eqref{fact}, as promised.
We finally put everything together in a similar fashion of \eqref{rep3}, to obtain:
 \begin{equation} \label{qrep3}
\langle V^n\rangle_{\xi} = \frac{
\int \left(\prod_{\rho=1}^n dE_\rho\right)\left(\prod_{\rho=1}^n \prod_{\sigma<\rho} dF_{\rho\sigma}dq_{\rho\sigma}\right) e^{NG\{q,F,E\}}
}{
\int \left(\prod_{\rho=1}^n dE_\rho\right) e^{NH\{E\}}
},
\end{equation}
where function $H\{E\}$ is identical the one we evaluated in the previous case \eqref{hfunc}, and function $G\{q,F,E\}$, albeit looking formally the same as \eqref{G}, hides a difference under the $K\{\lambda,x,q\}$ function \eqref{eqK}, this time quadratic instead of linear in $q_{\rho\sigma}$. 
\\ Assuming replica symmetry \eqref{replica}, this $K$ function becomes:
\begin{equation}
K\{\lambda,x,q\}=i \sum_{\rho=1}^nx_\rho\lambda_\rho-\frac{1-q}{2}\sum_{\rho=1}^nx_\rho^2-\frac{q^2}{2}\left( \sum_{\rho=1}^n x_\rho\right)^2.
\end{equation}
Once again we linearize the last term with the Gaussian integral trick:
\begin{equation}
e^{-\frac{q^2}{2}\left(\sum_\rho x_\rho\right)^2}=\int \frac{dt}{\sqrt{2\pi}}e^{-t^2+it{q}\sum_\rho x_\rho},
\end{equation}
and the integrals over the $x_\rho$'s and $\lambda_\rho$'s factorize:
\begin{equation}\label{eq130}\begin{aligned}
\int \left(\prod_{\rho=1}^n  \frac{dx_\rho\,d\lambda_\rho}{2\pi}\right) \,e^{K\{\lambda,x,q\}}&=
\int \frac{dt}{\sqrt{2\pi}} e^{-t^2}\prod_{\rho=1}^n\int  \frac{dx_\rho\,d\lambda_\rho}{2\pi} \, e^{i(\lambda_\rho+{q}t)x_\rho-\frac{1-q}{2}x_\rho^2}=\\
&=\int \frac{dt}{\sqrt{2\pi}} e^{-t^2}\left(\int  \frac{dx\,d\lambda}{2\pi} \, e^{i(\lambda+{q}t)x-\frac{1-q}{2}x^2}\right)^n=\\
&=\int \frac{dt}{\sqrt{2\pi}} e^{-t^2} \left(\int_\kappa^\infty\frac{d\lambda}{\sqrt{2\pi(1-q)}}\exp\left(-\frac{(\lambda+t{q})^2}{2(1-q)}\right)\right)^n.
\end{aligned}
\end{equation}
We stress once again that the only difference between eq. \eqref{eq130} and \eqref{eq119} lies in the behaviour of the $q$ variable. Taking the small $n$ limit of $G$, the first two terms in are unchanged, so they yield the exact same contribution of \eqref{eq121}, while the last one behaves like \eqref{eq120} (except for the $q$). Overall we have:
\begin{equation}\label{qg3}
\begin{aligned}
G(q,E,F)&=\frac{1}{2}n\left(E+qF+\log(2\pi)-\log(E+F)+\frac{F}{E+F}\right)+\\
&+ n\alpha \int \frac{dt}{\sqrt{2\pi}} e^{-t^2} \log\int_\kappa^\infty\frac{d\lambda}{\sqrt{2\pi(1-q)}}\exp\left(-\frac{(\lambda+t{q})^2}{2(1-q)}\right).
\end{aligned}
\end{equation}
Since the first two terms are unchanged, the two saddle-point equations $\frac{\partial G}{\partial E}=0$ and $\frac{\partial G}{\partial F}=0$ yields the same constraints as before, \eqref{eq122}. Putting these results into eq. \eqref{qg3} yields eq. \eqref{qg}, which we can then minimize; and taking the limit of dense solutions $q \to 1$ let us express $\alpha$ as a function of $\kappa$. Taking finally the limit $\kappa\to 0$ yields the desired result, $\alpha=4$.

\newpage

\printbibliography

\end{document}